\documentclass[lettersize,journal]{IEEEtran}
\usepackage{amsmath,amsfonts}
\usepackage{algorithmic}
\usepackage{algorithm}
\usepackage{array}
\usepackage[caption=false,font=footnotesize,labelfont=rm,textfont=rm]{subfig}
\usepackage{textcomp}
\usepackage{stfloats}
\usepackage{url}
\usepackage{amsmath}
\usepackage{tabularx}
\usepackage{verbatim}
\usepackage{graphicx}
\usepackage{amsmath}
\usepackage{amsfonts}
\usepackage[table]{xcolor}
\usepackage{soul, color, xcolor}
\usepackage{gensymb}
\usepackage{booktabs}
\usepackage{bbding}
\usepackage{bm}
\usepackage{multirow}
\usepackage{natbib}
\usepackage[utf8]{inputenc}
\usepackage[greek,russian,spanish,english]{babel}
\usepackage[OT2,T1, T2A,OT1]{fontenc}
\usepackage{graphicx, subfig}
\usepackage{utfsym}
\usepackage[inline]{enumitem}
\usepackage{makecell}
\usepackage{caption}
\usepackage{graphicx}
\usepackage{float} 
\usepackage{subfloat}
\usepackage{subcaption}
\usepackage{threeparttable}
\newtheorem{remark}{Remark}
\setcitestyle{numbers,square}
\usepackage[
pdfauthor={derajan},
pdftitle={How to do this},
pdfstartview=XYZ,
bookmarks=true,
colorlinks=true,
linkcolor=blue,
urlcolor=blue,
citecolor=blue,
pdftex,
bookmarks=true,
linktocpage=true,   
hyperindex=true
]{hyperref}
\usepackage{algorithm}
\usepackage{algorithmic}

\makeatletter
\newcommand{\removelatexerror}{\let\@latex@error\@gobble}
\makeatother

\begin{document}
	\title{Multiple Source Localization via Local Radio Map Construction in Urban Environments}
	
	\author{Qilu~Zhang,~Hongying~Tang*,~Wen~Chen,~\IEEEmembership{Senior Member,~IEEE},~Ziyi~Song and~Jiang~Wang*
		\thanks{This work was supported by NSFC under Grant 62531015 and U25A20399, and Shanghai Kewei under Grant 24DP1500500. \textit{(Corresponding authors: Hongying Tang and Jiang Wang.)}}
		\thanks{Qilu Zhang and Ziyi Song are with the Science and Technology on Micro-system Laboratory, Shanghai Institute of Microsystem and Information Technology, Chinese Academy of Sciences, Shanghai 200050, China, and also with the University of Chinese Academy of Sciences, Beijing 100049, China (e-mail: zhangqilu22@mails.ucas.ac.cn; songziyi22@mails.ucas.ac.cn).}
		\thanks{Hongying Tang and Jiang Wang are with the Science and Technology on Micro-system Laboratory, Shanghai Institute of Microsystem and Information Technology, Chinese Academy of Sciences, Shanghai 200050, China (e-mail: tanghy@mail.sim.ac.cn; jiang.wang@mail.sim.ac.cn).}
		\thanks{Wen Chen is with the Department of Electronic Engineering, Shanghai Jiao Tong University, Shanghai 200240, China (e-mail: wenchen@sjtu.edu.cn).}
		
	}
	
	
	\IEEEpubid{}
	
	\maketitle
	
	\begin{abstract}
		Accurately and efficiently addressing the multiple source localization (MSL) problem in urban environments, particularly designing a general method adaptable to an arbitrary number of sources, plays a crucial role in various fields such as cognitive radio (CR).
		Existing methods either fail to effectively utilize received signal strength (RSS) information without redundancy or lack generalizability to an arbitrary number of sources.
		In this work, we propose the Local Radio Map-Aided Multiple Source Localization Framework (LRM-MSL), which is a general method capable of handling an arbitrary number of sources.
		First, this framework constructs a local radio map that retains only the RSS information around the sources and binarizes it. Then, the connected component analysis tool is applied to the binarized map, which implements multi-source separation, transforming the MSL problem into a series of single-source localization (SSL) tasks.
		Finally, we design a numerical coordinate regression network to perform the SSL tasks.
		Since there is no publicly available RSS dataset for MSL, we construct the VaryTxLoc dataset to evaluate the performance of LRM-MSL. Experimental results demonstrate that LRM-MSL is an accurate and effective method, outperforming state-of-the-art approaches. Our code and dataset can be downloaded from https://github.com/hereis77/LRM-MSL.
		
	\end{abstract}
	
	\begin{IEEEkeywords}
		Deep learning (DL), radio map, multiple source localization (MSL), urban environment, received signal strength (RSS).
	\end{IEEEkeywords}

	\section{Introduction} \label{introduction}
	\IEEEPARstart{S}{ource} localization plays a critical role in cognitive radio (CR) networks, particularly in applications such as interference mitigation \cite{9894071}, detection of unauthorized sources \cite{LAZNA2024}, dynamic spectrum sharing, and enhancement of spectrum efficiency \cite{7942139}. In these contexts, accurate localization is essential for identifying primary users, avoiding harmful interference, and optimizing spectrum utilization to support reliable communication in dynamic environments.
	Typically, source localization is achieved by utilizing measurements collected by sensors deployed within the Area of Interest (AoI). These data serve as localization parameters, which can include Time of Arrival (ToA) \cite{guvenc2009survey}, \cite{shen2012accurate}, \cite{10747182}, Direction of Arrival (DoA) \cite{10897748}, Time Difference of Arrival (TDoA) \cite{ho2012bias}, \cite{zou2020tdoa}, Angle of Arrival (AoA) \cite{wang2015asymptotically}, \cite{7120424}, and Received Signal Strength (RSS) \cite{6638425}, \cite{7762904}, among others. Among these, RSS has garnered significant attention due to its cost-effectiveness and simple hardware requirements \cite{1458287}, and is thus chosen as the localization parameter in our work. 

	The advancement of emerging technologies, such as the CR networks and the fifth-generation mobile communication technology, have led to an exponential increase in the number of wireless devices. The surge in devices creates a challenge as multiple sources transmit simultaneously in AoI, leading to signal overlap. This requires localization methods to handle mixed RSS values of multiple sources and accurately localize them, whose positions and numbers are typically unknown.

	\begin{figure}[t]
		\centering
		\includegraphics[width=0.48\textwidth]{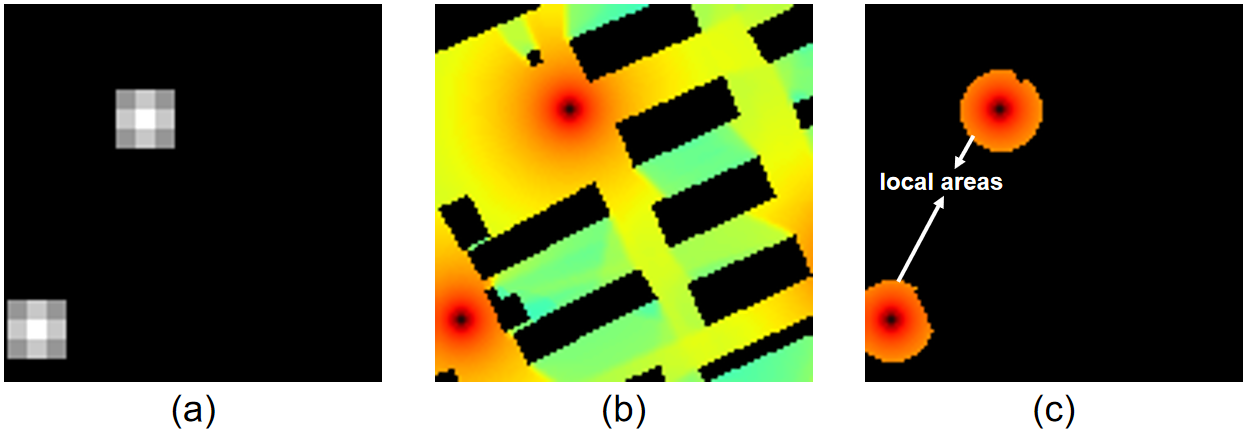}
		\caption{Existing intermediate maps can be: (a) Heatmap, where the vicinity of sources does not contain RSS information. (b) Global radio map, which contains all the RSS information of the AoI. (c) Local radio map, which retains only the RSS information of the local area.}
		\label{global_local}
		\vspace{-0.5cm}
	\end{figure}

	Recently, deep learning (DL)-based methods have gained widespread attention due to their ability to learn representations of data with different levels of abstraction as well as complex and non-linear relationships from data \cite{lecun2015deep}, \cite{10409292}. This capability makes it a promising solution to the challenge of signal overlap in MSL.

	Traditionally, DL-based MSL methods first estimate the number of sources as prior information based on measurements. Based on this, a convolutional neural network (CNN) is designed to extract features from the measurements, with its output layer typically being a fully connected (FC) layer whose dimensions match the number of sources, mapping to the final coordinates. Bizon et al. \cite{bizon2023blind} adjust the output dimension of the FC layer and loss function based on the previously estimated number of sources and train the network for that specific number of sources.
	This method, however, can only handle scenarios with a fixed number of sources. Every time when the number of sources varies, the CNN needs to be retrained. 
	In order to deal with different number of sources, DeepTxFinder \cite{zubow2020deeptxfinder} establishes a network library containing \(N\) pretrained convolutional neural networks (CNNs), where \(N\) is the maximum number of sources expected in the AoI. It then selects the network corresponding to the estimated number of sources for localization.
	However, this approach has notable drawbacks. Specifically, a considerable amount of samples is required to train the CNN for each possible number of sources, leading to prolonged training time and significantly increasing the dataset size requirements. Moreover, the model is difficult to generalize to scenarios where the number of sources exceeds \( N \).

	To avoid training multiple CNNs for various number of sources, map-based methods have been proposed in \cite{zhan2021deepmtl}, \cite{mitchell2022deep}, \cite{10246277}, which consists of two stages: the measurements-to-map translation stage, where the measurements are mapped onto an intermediate map, and coordinate determination stage, where the sources’ coordinates are determined by employing an image-based peak-finding operation. In this way, map-based methods only need to train a single CNN using samples with various numbers of sources, thereby enhancing the model's generalization capability.

	In the measurements-to-map translation stage, an intermediate map is first constructed using the measurements. There are two types of existing intermediate maps: heatmap and global radio map. 
	First, the heatmap is an artificially designed map, typically representing the AoI as a normalized grayscale image. In this map, the vicinity of a source is represented as a small patch with pixel values ranging from [0, 1], while pixels outside the patch are set to 0. As shown in Fig. \ref{global_local}(a), the center pixel is set to 1, and the surrounding pixel values within the patch smoothly decay as the distance from the source increases. DeepMTL \cite{zhan2021deepmtl} adopts gaussian distributions in the heatmap to represent the sources’ pixel patches. 
	Mitchell et al. \cite{mitchell2022deep} represent each source as a 3$\times$3 pixel patch, with a center value of 1 and surrounding values of 0.5. 
	Second, the global radio map is constructed by completing sparse measurements in radio communication environments, such as RSS, thereby reflecting the global distribution of electromagnetic energy, as shown in Fig. \ref{global_local}(b). Various techniques have been applied to reconstruct global radio maps, including interpolation methods like radial basis functions (RBF) \cite{7952827}, Kriging \cite{8455448}, and deep learning-based methods such as autoencoders (AE) \cite{9523765}, \cite{9354041}, \cite{10608081}, Graph Neural Networks (GNN) \cite{10078269}, Generative Adversarial Networks (GAN) \cite{10025551}, \cite{10130091}, \cite{10615679}, Diffusion Model \cite{10764739}, among others.
	Furthermore, several methods based on global radio maps have been proposed for MSL. Wu et al. \cite{10525070} utilize an AE to reconstruct the global radio map, while an approach proposed in \cite{10246277} uses a conditional generative adversarial network (cGAN) to complete the raw sparse measurements into a global radio map.  
	However, the existing types of these intermediate maps have inherent limitations: 1) The constructed heatmaps cannot fully exploit the RSS information around the sources, which can reduce localization accuracy. 2) The global radio maps contain redundant RSS information from distant areas, diluting the focus on critical RSS information surrounding the sources.

	In the coordinate determination stage, existing works \cite{zhan2021deepmtl}, \cite{mitchell2022deep}, \cite{10246277} determine the source location by identifying peak values on the map. In \cite{mitchell2022deep}, the sources’ coordinates are identified by applying a thresholding technique, classifying any pixel with a value greater than 0.2 as a potential source, along with a 3×3 filter to suppress detection near local maxima. This method only achieves pixel level localization.
	The approach proposed in \cite{10246277} adopts persistence calculation and thresholding, treating pixels with persistence above the threshold as peaks. Taking the contributions of neighboring pixels into account, the positions of the sources are refined to sub-pixel accuracy by calculating the weighted sum of the four-pixel neighborhood.
	However, it is merely a naive parameter calculation approach, considering only the influence of four neighboring pixels for localization, which may reduce accuracy.
	Considering the methods \cite{mitchell2022deep}, \cite{10246277} cannot capture the physical propagation relationships between positional changes and the corresponding RSS variations, DeepMTL \cite{zhan2021deepmtl} proposes a DL-based YOLOv3-custom object detector, which determines the anchor box, and the location of the sources is subsequently identified as the center of the anchor box. 
	Unfortunately, the object detector require additional anchor box labeling on the intermediate map and the knowledge of numerical coordinates of sources. As such, this process is both time-consuming and labor-intensive.

	Rencently, Wu et al. \cite{10525070} utilize FC layers to directly map the global radio map to coordinates, with the output dimension visually determined by the intermediate map. While methods using FC layers as the output layer can perform direct numerical coordinate regression without the need for complex label formats, they require retraining the model for different numbers of sources, thus limiting their generalizability to handle an arbitrary number of sources.
	
	With the above discussions, we summarize that two challenges in the existing MSL problems remain unresolved:
	
	\begin{enumerate}
		\item{Is there such an intermediate map that leverages the RSS information around sources while eliminating the influence of distant redundant RSS data?}
		\item{Is it possible for a single, general CNN with direct numerical coordinate supervision to effectively perform coordinate determination for an arbitrary number of sources?}
	\end{enumerate}
	
	To deal with these issues, in this paper, we introduce a novel intermediate map, the local radio map, which preserves the true RSS distribution only within a \textit{local area}, defined as the area centered at a source with radius r, while setting pixel values outside this area to zero, thereby removing redundant information far from the sources.
	As shown in Fig. \ref{global_local}(c), the RSS values exhibit a sharp drop between the local and non-local areas, creating a clear and fixed boundary that eliminates the complexity of threshold selection and ensures consistency in defining the local area for each source.
	
	To enable a single coordinate regression network to handle an arbitrary number of sources, we introduce multi-source separation in the measurements-to-map translation stage, transforming the MSL problem into multiple SSL problems. Specifically, we extract \(M\) local areas from the local radio map and generate \(M\) single-source local radio maps to achieve multi-source separation. 
	In the coordinate determination stage, we design a general numerical coordinate regression network, with each single-source local radio map serving as an input feature.
	This network employs an FC layer with a fixed dimension corresponding to a single source as the output layer, designed to solve the SSL problem. The network is trained to capture the physical propagation characteristics between position changes and RSS variations, which facilitates high-accuracy localization. On the other hand, its output dimension remains fixed regardless of the number of sources, ensuring general applicability to an arbitrary number of sources. 
	
	Furthermore, as urbanization advances, the demand for MSL in cities grows. Accurate localization is vital for applications such as tracking illegal sources, optimizing resources, and enhancing public safety. Therefore, considering the MSL methods in urban environments is both necessary and valuable. Drawing inspiration from the urban radio map reconstruction approach in \cite{9354041}, we take urban environments into consideration. 
	Our contributions can be summarized as follows:
	\begin{itemize}
		\item \textit{The concept of local radio map:}
		We introduce the concept of the Local Radio Map, which solely displays RSS variations within a fixed local area, helping the neural network focus on pixel variations relevant to localization and facilitating subsequent multi-source separation.
		
		\item \textit{A two-stage framework \textemdash Local Radio Map-Aided Multiple Source Localization Framework (LRM-MSL):} In the first stage, the local radio map is constructed by a neural network from sparse measurements and building layouts. Subsequently, multi-source separation is performed through binarization and connected component analysis, transforming the MSL problem into multiple SSL problems. In the second stage, SSL tasks are solved by a numerical coordinate regression network.
		\item \textit{Performance and robustness:}
		We conduct extensive experiments to demonstrate that performance of LRM-MSL outperforms other state-of-the-art methods. In addition, ablation studies verify the effectiveness of the overall framework design.
		
	\end{itemize}

	To provide a thorough understanding of our approach, we organize the remainder of the paper as follows: Section \ref{Related Work} introduces the related works. In Section \ref{Preliminaries}, we present the system model and the foundational concepts relevant to our work. Section \ref{Multiple source localization with deep learning} focuses on multiple source localization using deep learning, providing a comprehensive analysis of our approach. We then demonstrate the explanations of numerical experiments in Section \ref{Performance Evaluation and Results}, highlighting the effectiveness of our method. Finally, we conclude the paper in Section \ref{Conclusion}.

	\begin{figure}[t]
		\centering
		\includegraphics[width=0.36\textwidth]{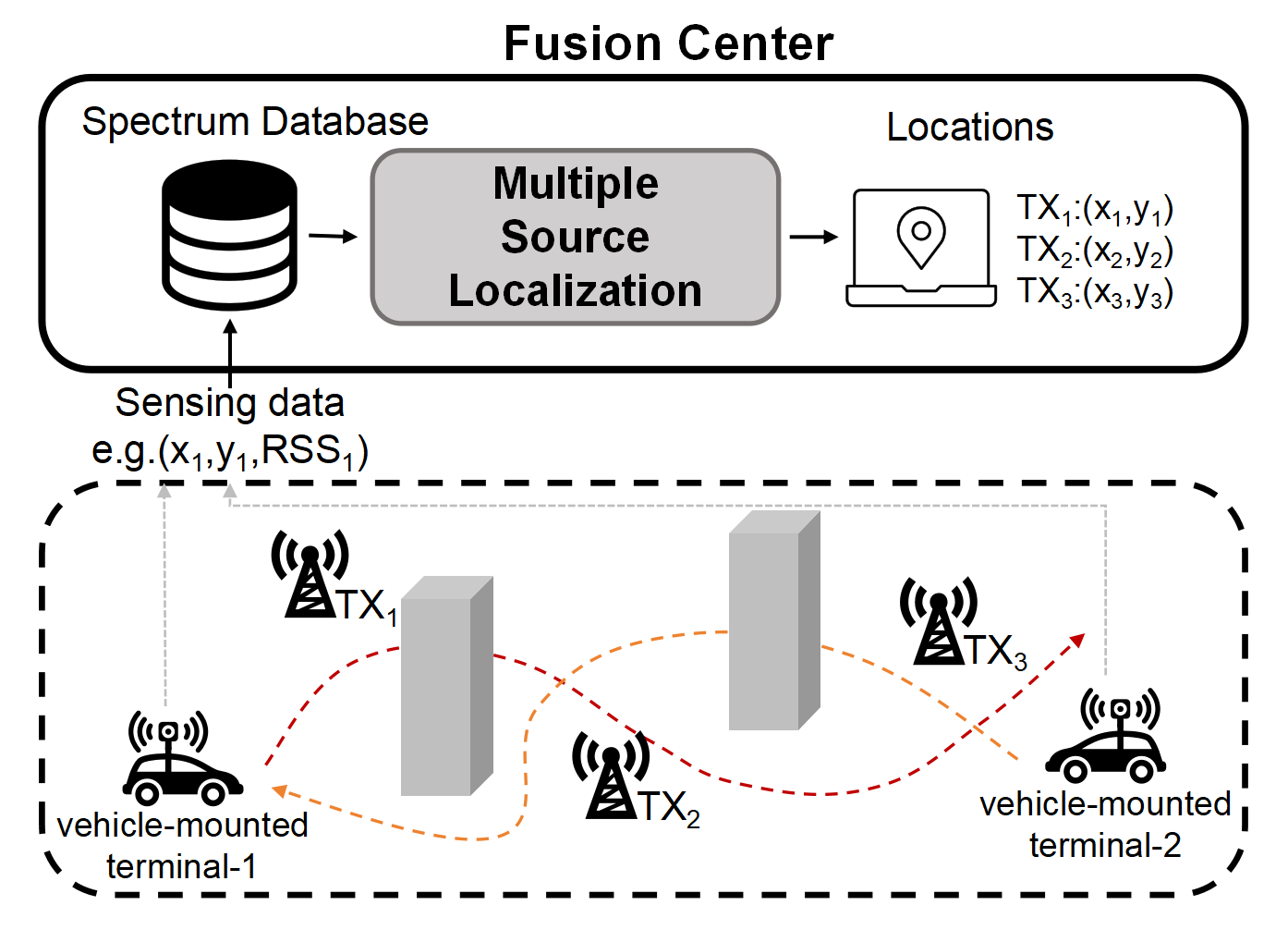}
		\caption{Spectrum monitoring and MSL
			framework. Vehicle-mounted terminals collect RSS measurements
			from multiple sources, which is processed at the fusion
			center using a spectrum database to estimate the locations of
			the sources.}
		\label{framework}
		\vspace{-0.5cm}
	\end{figure}

	\section{Related Work} \label{Related Work}
	
	\subsection{Model-based RSS Localization Methods}
	Model-based methods rely on physical or mathematical models, where prior modeling of signal propagation laws and environmental characteristics is used to solve the source position.
	The key idea is to establish an analytical relationship between the observations and the source location based on known model parameters, and then infer the position through mathematical derivation or optimization algorithms.
	Typical representatives include recursive Bayesian estimation, maximum likelihood estimation, and compressive sensing.
	Wang et al.~\cite{8532443} converted the log-normal shadowing model into a multiplicative form and applied semidefinite programming (SDP) for cooperative localization.
	Shi et al.~\cite{9127148} addressed unknown transmit power via a least-squared relative error estimator with SDP relaxation.
	Ziskind and Wax~\cite{7543} proposed an alternating projection maximum likelihood method for multi-source localization.
	Sun et al.~\cite{9372898} developed a robust weighted least squares approach for mixed LOS/NLOS environments, jointly estimating source position and unknown parameters using SDP relaxation.
	Jiang et al.~\cite{9268105} applied Bayesian compressive sensing with UAV-assisted RSS measurements, optimizing UAV trajectories via adaptive simulated annealing for multi-source localization.
	These methods typically require certain prior information, such as the path-loss exponent, shadow fading statistics or reference distance.
	They offer interpretability in the computation process but their performance heavily depends on the match between the assumed model and the actual scenario.
	When the environment is complex or model parameters are unknown, particularly in urban environments, the localization accuracy can be significantly affected.

	\subsection{Model-free RSS Localization Methods}
	In model-free RSS localization methods, localization relies solely on RSS measurements and their corresponding positions. The most common approach is fingerprint-based localization \cite{10835800}, where an offline database maps RSS patterns to known locations. During online localization, new measurements are matched to the database to estimate source positions. While effective, fingerprinting is resource-intensive and highly sensitive to environmental changes, motivating recent interest in deep learning (DL) solutions.
	
	Many DL-based methods employ CNNs with fully connected layers to directly regress source coordinates, usually after first estimating the number of sources. Bizon et al. \cite{bizon2023blind} adapt the FC layer and loss for each source count, while DeepTxFinder \cite{zubow2020deeptxfinder} maintains a library of CNNs for all possible source numbers. Other approaches map RSS data into intermediate maps. For example, DeepMTL \cite{zhan2021deepmtl} and Mitchell et al. \cite{mitchell2022deep} convert measurements into heatmaps, representing local regions with Gaussian kernels or fixed pixel patches, and then apply object detection or thresholding to identify sources. Wu et al. \cite{10525070} reconstruct global radio maps via an AE and regress coordinates with FC layers. Similarly, \cite{10246277} and \cite{10829585} complete sparse measurements into global maps using a cGAN and IMNet, respectively, followed by threshold-based heuristics for localization. 
	Overall, existing DL approaches either suffer from poor scalability, requiring retraining when the source number changes, or struggle to fully exploit surrounding RSS information and minimize redundancy in global radio map representations, limiting localization accuracy.

	\section{System Model} \label{Preliminaries}
	We consider a system as illustrated in Fig.~\ref{framework}, where a fusion center is responsible for multi-source localization (MSL). The term \textit{multi-source} refers to multiple sources operating within the same communication system. We assume that a spectrum sensing procedure has been carried out prior to the MSL task to identify the target frequency point and bandwidth. Vehicle-mounted terminals then collect RSS measurements from these sources operating within the selected frequency band, and transmit the measurements to the fusion center. The center subsequently processes the RSS measurements to accurately estimate the locations of these sources.
	Consider an AoI \(\mathcal X \subset \mathbb{R}^2\), where \textit{M} sources with unknown locations are placed at fixed locations. We denote their 2D locations by \(\textbf{Q}= [Q_1, Q_2, ..., Q_M]\) where \(Q_i = [x_i,y_i]\). We assume a stationary signal emitted by a source for a period long enough to be identified by spectrum sensing. A certain number of vehicle-mounted terminals collect RSS measurements\(\{\widetilde{\Psi}(S_j)\}_{j=1}^J\) at $J$ locations \(\textbf{S}= [S_1, S_2, ..., S_J]\) where  \(S_j=[x_j,y_j]\).
	
	Notably, sensing devices cannot separate the mixed signal and instead observe the aggregated signal strength from all sources.
	Assume that the RSS at a given location $S_j$ within the AoI corresponding to the $i$-th source is denoted by $\Psi_{\text{RX}}^{i,j}$. This can be expressed as:
	\begin{equation}
		\Psi_{\text{RX}}^{i,j} = \Psi_{\text{TX}}^i - L(d)^{i,j} - \xi^{i,j} - \zeta^{i,j}, \label{RSS}
	\end{equation}
	where $\Psi_{\text{TX}}^i$ represents the transmission power of $i$-th source, $L(d)^{i,j}$ denotes the path loss in free space, $\xi^{i,j}$ models the shadowing component, and these two components jointly represent large-scale fading. $\zeta^{i,j}$ represents the small-scale fading, which is customarily assumed to have been averaged out \cite{9523765}, \cite{10103465}, \cite{10682525}.

	Specifically, the main factors considered are path loss and shadowing. The path loss $L(d)^{i,j}$ is typically modeled as a log-normal random variable, commonly expressed as
	\begin{equation}
		L(d)^{i,j} = L_0 + 10\eta\log_{10}\left(\frac{d^{i,j}}{d_0}\right) + \chi_\sigma,
	\label{pathloss}
	\end{equation}
	where $L_0$ is the path loss at a reference distance $d_0$, $\eta$ is the path loss exponent, $d^{i,j}$ is the distance between the $i$-th source and the location $S_j$, and $\chi_\sigma$ is a Gaussian random variable with zero mean and variance $\sigma^2$~\cite{10608081}. The shadowing component $\xi^{i,j}$ is also modeled as a zero-mean Gaussian random variable, i.e., $\xi^{i,j} \sim \mathcal{N}(0, \sigma_{\xi^{i,j}}^2)$~\cite{molisch2012wireless}.
	
	The total RSS received at $S_j$ is equal to
	\begin{equation}
		\widetilde{\Psi}(S_j) = \sum_{i=1}^M\Psi_{\text{RX}}^{i,j}.
	\end{equation}
	
	Moreover, we assume that each vehicle-mounted terminal is able to measure its own real-time location with high accuracy and convey it to the fusion center. Given \(\{(S_j,\widetilde{\Psi}(S_j))\}_{j=1}^J\), the problem that the fusion center needs to solve is to predict all unknown sources coordinates based solely on observed sparse RSS measurements and building layout.

	\begin{figure*}[t]
		\centering
		\includegraphics[width=1\textwidth]{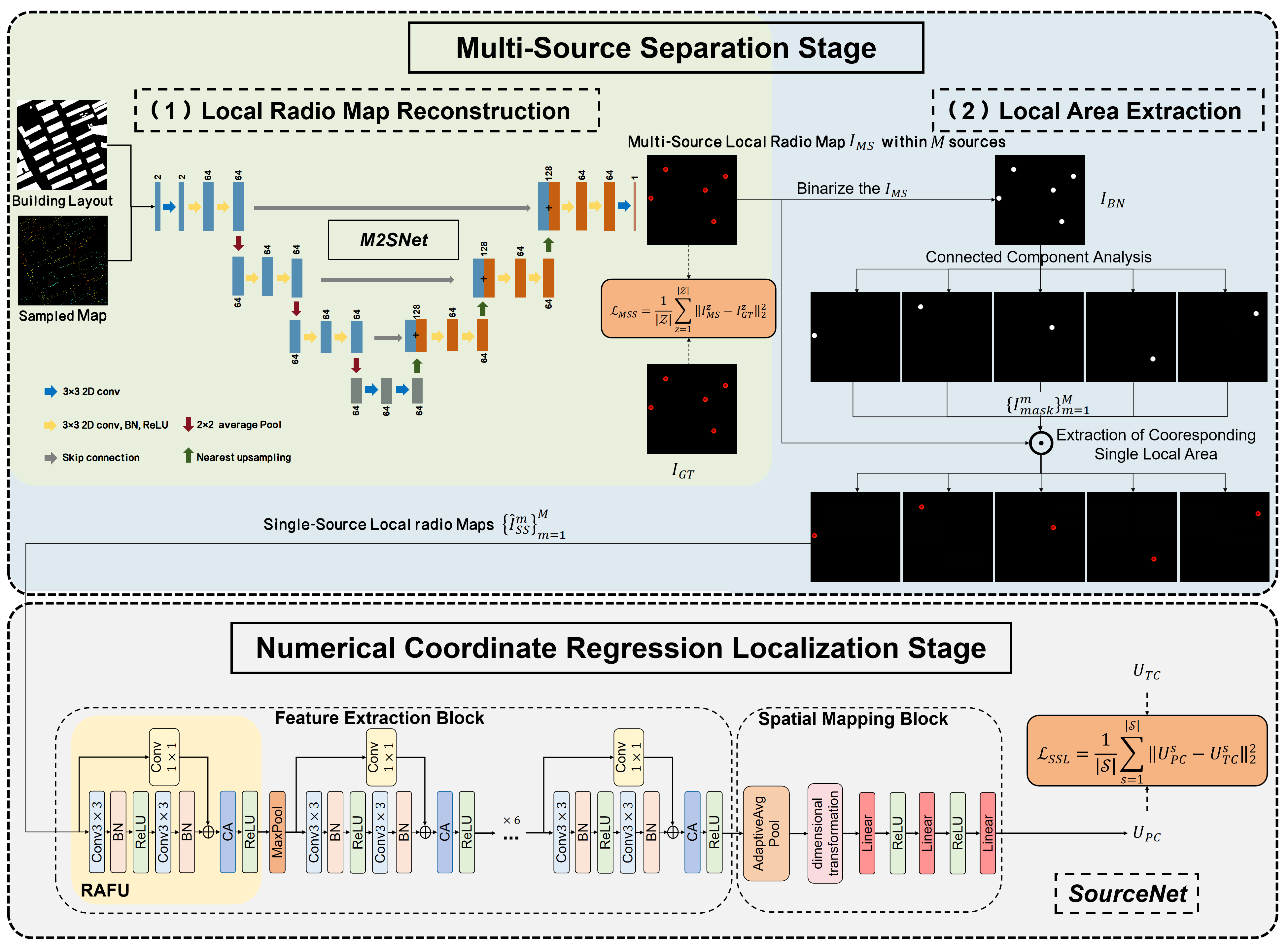}
		\caption{Overall Structure of LRM-MSL.}
		\label{overview}
		\vspace{-0.5cm}
	\end{figure*}

	\section{Multiple source localization with deep learning}   \label{Multiple source localization with deep learning}
	\label{method}
	This section introduces a novel two-stage approach for localizing multiple unknown sources, which is implemented through our proposed Local Radio Map-Aided Multiple Source Localization Framework (LRM-MSL).
	The two stages consist of: multi-source separation and numerical coordinate regression localization. 
	In the first stage, the multi-source local radio map \(I_{MS}\) is reconstructed, which is a bitmap ranging from 0 to 255 and displays the RSS distribution of \(M\) sources within the AoI. 
	Then, \( M \) local areas are extracted from the \(I_{MS}\), generating \( M \) single-source local radio maps \(\{\hat{I}_{SS}^m\}_{m=1}^M\), achieving multi-source separation and facilitating the transformation of the MSL problem into multiple SSL tasks.
	In the second stage, SourceNet is designed based on numerical coordinate regression network to perform SSL tasks, predicting the coordinates of individual sources using the seperated connected components from the previous stage.
	The approach's overview is depicted in Fig. \ref{overview}.

	\subsection{Multi-Source Separation} 
	The multi-source separation stage consists of two key parts: \textbf{local radio map reconstruction} and \textbf{local area extraction}.
	The former lays a foundation for multi-source separation, facilitating transforming MSL problems into SSL problems.

	The first part, \textbf{local radio map reconstruction}, utilizes an M2SNet to reconstruct the \(I_{MS}\) based on the input features: 1) the sampling map and 2) the building layout (cf. Section \ref{Dataset}). M2SNet is composed of multiple encoding and decoding stages, with each stage connected by skip connections that help preserve spatial information. The encoder progressively downsamples the input through convolutional layers, batch normalization (BN), and ReLU activations, followed by pooling operations that reduce the spatial dimensions while capturing essential features. After passing through the bottleneck layer, which compresses the learned representations, the decoder begins the upsampling process, gradually restoring the feature maps to their original resolution. Skip connections are used to link the corresponding feature maps from the encoder, allowing the model to retain high-resolution details that are crucial for precise reconstruction. The overall architecture is shown in Fig. \ref{overview}. Although we use M2SNet, which adopts a simple encoder-decoder structure in this work, we believe that other networks could perform similarly for this task, and our approach is not intended to propose a novel network architecture.
	
	The final output $I_{MS}$ and ground truth $I_{GT}$ of the M2SNet are both images. We employ the L2 loss function to calculate the MSE of individual pixels as a metric to evaluate the accuracy of the proposed method on a given subset $\mathcal{Z}$ of the dataset. More specifically, our loss function is defined as:
	\begin{equation} 
		\mathcal{L}_{MSS}=\frac{1}{\mathcal{\lvert Z \rvert}}\sum_{z=1}^\mathcal{\lvert Z \rvert} \|I_{MS}^z - I_{GT}^z \|_2^2,
	\end{equation}
	where \(I_{MS}^z\) and \(I_{GT}^z\)  denote the M2SNet's output and the corresponding ground truth of the $z$th instance of the dataset. 
	
	\begin{figure}[!t]
	
	\removelatexerror
	\begin{algorithm}[H]
		\caption{Local Area Extraction}
		\label{alg:connected_component}
		\begin{algorithmic}[1]
			\REQUIRE An \(M\)-source multi-source local radio map \(I_{MS}\) 
			\ENSURE A set of single-source local radio maps \(\{\hat{I}_{SS}^m\}_{m=1}^M\) 
			\FOR{each pixel \( (i,j) \)}
			\STATE  \textbf{\( I_{MS}(i,j) \in [0, 255] \)}
			\STATE \( \gamma \gets 127 \) 
			\vspace{-2mm} 
			\STATE \( I_{BN}(i,j) \gets \begin{cases} 
				255 & \text{if } I_{MS}(i,j) > \gamma \\ 
				0 & \text{if } I_{MS}(i,j) \leq \gamma    
			\end{cases} \)  
			\ENDFOR
			
			\STATE \( \{ I_{mask}^m \}_{m=1}^M \gets \text{connectedComponents}(I_{BN}) \) \textbf{(OpenCV function for connected component analysis)}
			\STATE \textbf{Normalize} \( \{I_{mask}^m \}_{m=1}^M \) to \( [0, 1] \)

			\FOR{\( m \) from 1 to \( M \)}
			\STATE \( \hat{I}_{SS}^m(i,j) \gets I_{mask}^m(i,j) \odot I_{MS}(i,j) \)
			\ENDFOR
			\STATE \textbf{Return} \(\{\hat{I}_{SS}^m\}_{m=1}^M\)

		\end{algorithmic}
	\end{algorithm}
	\vspace{-0.5cm}
	\end{figure}
	
	The second part of this stage, termed \textbf{local area extraction}, results in the generation of distinct \(\{\hat{I}_{SS}^m\}_{m=1}^M\). The input image of this part is \( I_{MS} \), which retains the information within a radius \( r \) centered on the sources, while the surrounding areas are set to zero. 
	We first binarize the \( I_{MS} \), which has a pixel range of 0 to 255, using a threshold of \( \gamma = 127 \) to obtain the binarized map \( I_{BN} \), where the local area is set to 255 while the rest is set to 0, with no intermediate grayscale levels (See Remark \ref{Remark1}). In image processing, regions of adjacent foreground pixels with the same pixel value are typically referred to as a connected component. Therefore, the local area in \( I_{BN} \) is considered a connected component.
	Then, we perform connected component analysis\footnote{For detailed information on connected component analysis, refer to the OpenCV documentation: \url{https://docs.opencv.org/3.4/d3/dc0/group__imgproc__shape.html}.}, a technique used to label and separate different connected components within the image, to identify the connected components of the \(M\) sources’ local areas in \(I_{BN}\), denoted as \(\{I_{mask}^m\}_{m=1}^M\). Each connected component corresponds to a local area, which contains a source. 
	Following this, the binary mask \( I_{mask}^m \) is element-wise multiplied with the \( I_{MS} \) to retain the RSS information around the $m$th source, while zeroing out the rest of the image. This operation ensures that the extracted single-source local radio map \( \hat{I}_{SS}^m \) is aligned with the relevant area in \( I_{MS} \). The implementation details of the local area extraction is provided in Algorithm \ref{alg:connected_component}.

\begin{figure}[t]
		\centering
		\includegraphics[width=0.25\textwidth]{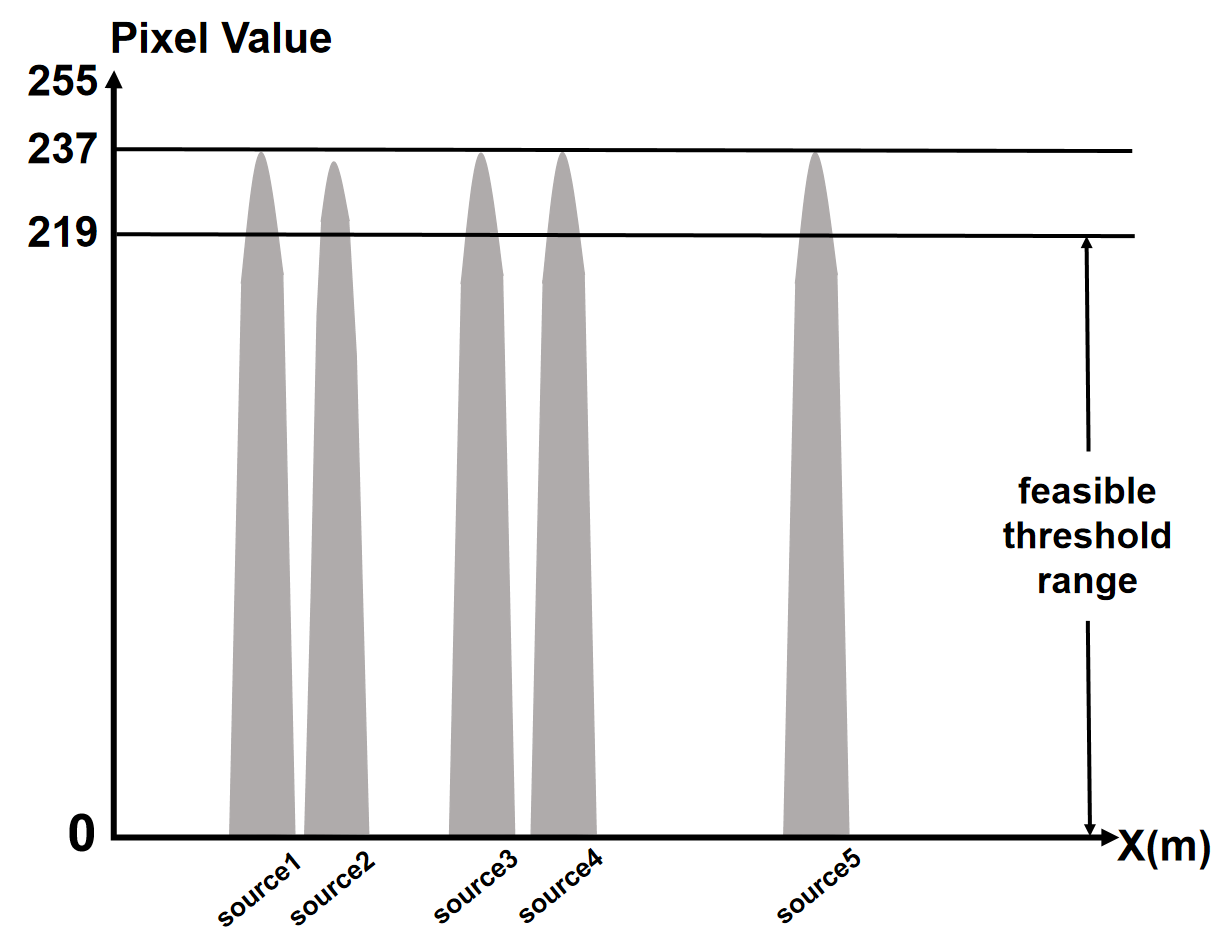}
		\caption{Pixel value variations reflecting RSS changes along the x-axis in the local radio map.}
		\label{2Dglobal_local}
		\vspace{-0.5cm}
	\end{figure}

	\begin{remark}
	\label{Remark1}
	\textit{To better illustrate the selection of the threshold \(\gamma\) for binarization, we showcase a local radio map along the x-axis when five sources transmit signals simultaneously.
	As shown in Fig. \ref{2Dglobal_local}, in this local radio map, the pixel values within the local areas follows in  range of [219, 237], while the others are zero, which exhibits a distinct boundary between the local and non-local areas. 
	We also observe from the results in the dataset (cf. \ref{Dataset}) used for the experiments that the pixel values of the local areas in each constructed local radio map exhibit slight fluctuations but remain within the range of [200, 255]. Therefore, using \(\gamma = 127\) as the binarization threshold in Algorithm 1 is reasonable.
	}
	\end{remark}

	\subsection{Numerical Coordinate Regression Localization} 
	In this section, we present the second stage of LRM-MSL. The purpose of this stage is to obtain numerical coordinates through the single-source local radio maps output in the previous stage, so as to achieve subpixel level localization. 
	
	We refer to our numerical coordinate regression localization model as \textit{SourceNet}. This is a coordinate regression network inspired by ResNet's residual learning strategy, which helps alleviate gradient vanishing in deep networks.
	As shown in Fig. \ref{overview}, the proposed SourceNet consists of two parts: the feature extraction block and the spatial mapping block. 
	To better demonstrate the model, we define $U_{PC}$ and $U_{TC}$ as the predicted coordinates, and the ground-truth coordinates.

\begin{figure}[t]
		\centering
		\includegraphics[width=0.245\textwidth]{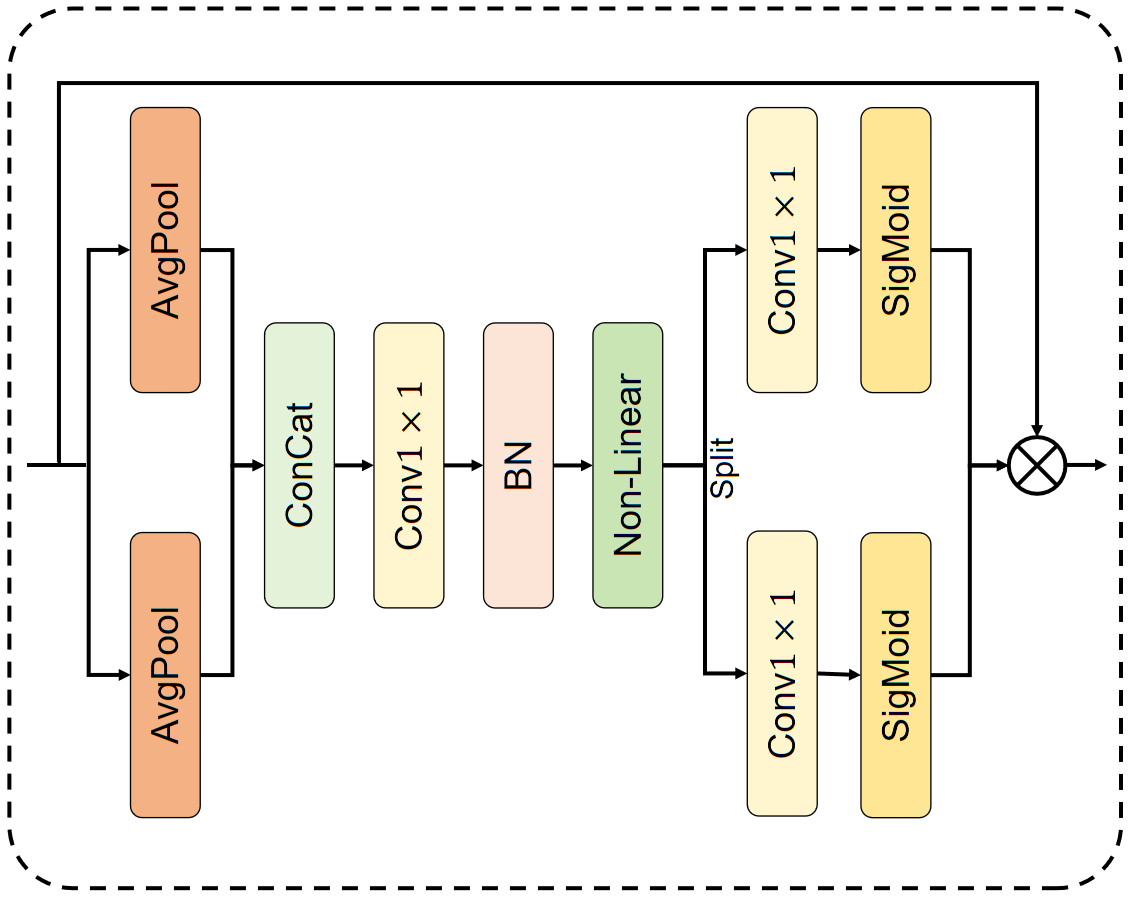}
		\caption{The coordinate attention (CA) module.}
		\label{ca}
		\vspace{-0.5cm}
	\end{figure}

	In the feature extraction block, we introduce a novel module, called the \textit{Residual Attention Feature Unit} (RAFU). 
	The network takes the single-source local radio maps and progressively learns important features from it through $6$ RAFUs connected via max pooling layers.
	Each RAFU includes two convolutional blocks, each of which incorporates a 3$\times$3 convolutional layer, a BN layer, and a ReLU activation function. Within the second convolutional block, to alleviate the problem of vanishing gradients, we also add the residual connection between the input of the RAFU and the output of BN(match dimensions by 1$\times$1 convolutional).
	Subsequently, to capture long-range dependencies in one spatial dimension while preserving precise positional information in the other, we introduce a Coordinate Attention (CA) \cite{hou2021coordinate} mechanism, as shown in Fig. \ref{ca}. The CA module enhances the network's ability to capture cross-channel dependencies and spatially relevant, direction-aware features, which are crucial for accurate source localization. Its ability to encode position-sensitive information motivates its integration into SourceNet, enabling the model to make more spatially informed decisions.
	The CA module received the combined output from the residual connection, followed by a BN layer for normalizing the feature maps, facilitating their further processing.
	
	The spatial mapping block maps the high-dimensional and complex feature space into a lower-dimensional space that is more suitable for the coordinate prediction task. To accomplish this, we feed the output of the last RAFU through an adaptive average pooling layer and a dimensional transformation layer, followed by a three-layer multi-layer perceptron comprising an input layer, a hidden layer, and an output layer. The input layer's neurons match the transformed features from the last RAFU. The output FC layer is tasked with generating the final source localization result, with a neuron number of $2$.

	The final output of SourceNet is the predicted coordinates $U_{PC}$, the accuracy of the proposed method over a given subset $\mathcal{S}$ of the dataset is measured by the average squared euclidean distance, i.e., MSE, between $U_{PC}$ and $U_{TC}$, given by
	
	\begin{equation} 
		\mathcal{L}_{SSL}=\frac{1}{\mathcal{\lvert S \rvert}}\sum_{s=1}^\mathcal{\lvert S \rvert} \|U_{PC}^s - U_{TC}^s \|_2^2,
	\end{equation}
	where $U_{PC}^s:=(x_{pc}^s,y_{pc}^s)$ and $U_{TC}^s:=(x_{tc}^s,y_{tc}^s)$ denote the SourceNet prediction and the corresponding true coordinates of the $s$th instance of the dataset.

	\section{Performance Evalution and Results}   \label{Performance Evaluation and Results}
	\subsection{The VaryTxLoc Dataset}  \label{Dataset}
	We construct VaryTxLoc, a dataset of building layouts with corresponding single-source and multi-source simulated radio maps, which are computed using the Intelligent Ray Tracing (IRT) mode of the WinProp software \cite{hoppe2017wave} \cite{rautiainen2002verifying}.
	
	1) \textit{Simulation Parameters:} 
	The simulation parameters in the VaryTxLoc dataset are selected based on established practices and previous studies \cite{9354041}, \cite{10078269}, as reported in Table \ref{para}. According to the classic single-source radio map dataset, RadioMapSeer \cite{9354041}, which has been widely used for radio propagation modeling, the heights of the sources and receivers are set to 1.5 m in our work. The building heights of 25 m also follow the RadioMapSeer, ensuring consistency with typical dataset in electromagnetic propagation studies.
	The center carrier frequency, bandwidth and transmit power are set to 2.0 GHz, 10 MHz and 24 dBm, respectively, referencing the 5.9 GHz, 10 MHz and 23 dBm from \cite{9354041} and 1.5 GHz, 10 MHz and 43 dBm from \cite{10078269}. 
	It is important to note that RadioMapSeer primarily focuses on path loss and is designed for single-source scenarios, while the VaryTxLoc uses RSS for MSL.
	
	2) \textit{Building Layout:} 
	The building layouts are taken from OpenStreetMap \footnote{\url{https://www.openstreetmap.org/}.}. Each building layout is \( 200 \times 200 \, \text{m}^2 \) where buildings are saved in the dataset as polygons. Each building layout is also converted to morphological 2D image (binary 0/1 pixel values with no intermediate gray levels) of $200 \times 200$ pixels, where each pixel represents one square meter. The interior of the buildings are set to a pixel value of 1, and otherwise set to 0.

	3) \textit{Global Radio Map:} 
	In this dataset, the global radio maps are divided into two subsets: single-source global radio maps and multi-source global radio maps. The single-source subset includes 24 different building layouts, each with 40 distinct source placements. 
	The multi-source subset also contains 24 building layouts, with scenarios involving 1, 3, 5, and 7 sources, each having 10 different placement layouts. Additionally, data augmentation techniques such as vertical and horizontal flipping, as well as 90°, 180°, and 270° rotations, were applied. As a result, the single-source and multi-source global radio map subsets each contain 5760 samples.

	4) \textit{Local Radio Map:} 
	The local radio maps are also divided into two subsets: single-source local radio maps and multi-source local radio maps. These subsets are derived by retaining only the pixels within a circular area of radius \( r \) centered at the coordinates of each source in the corresponding global radio map. All pixels outside these circular areas are set to zero. Similar to the global radio maps, the single-source and multi-source local radio map subsets each contain 5760 samples. The selection of radius \( r \) is based on the assumption that the minimum distance between two sources in a local radio map should be \( 2r \). 
	Specifically, we set \( r = 2 \), which implies a minimum distance of 4 m between two sources. This source layout condition is already considered highly dense within a \( 200 \times 200 \, \text{m}^2 \) area.
	
	5) \textit{Sampling Strategy and Sampled Map:} 
	Currently, most studies adopt uniform random sampling. Notably, during the actual deployment process, this sampling method presents two pivotal challenges. Firstly, the limited availability of hardware devices imposes constraints on the quantity and distribution range of sampling points. Under the uniformly random sampling approach, attaining sufficient sampling data necessitates a considerable number of hardware devices operating concurrently, which is impractical in real-world scenarios. Secondly, stringent conditions govern the distribution density of sampling points. Uniform random sampling necessitates an even distribution of sampling points across the entire AoI. However, due to limitations posed by terrain, environmental factors, and the deployment locations of hardware devices, achieving an ideal uniform distribution becomes exceedingly difficult.
	
\begin{figure}[t]
		\centering
		\includegraphics[width=0.3\textwidth]{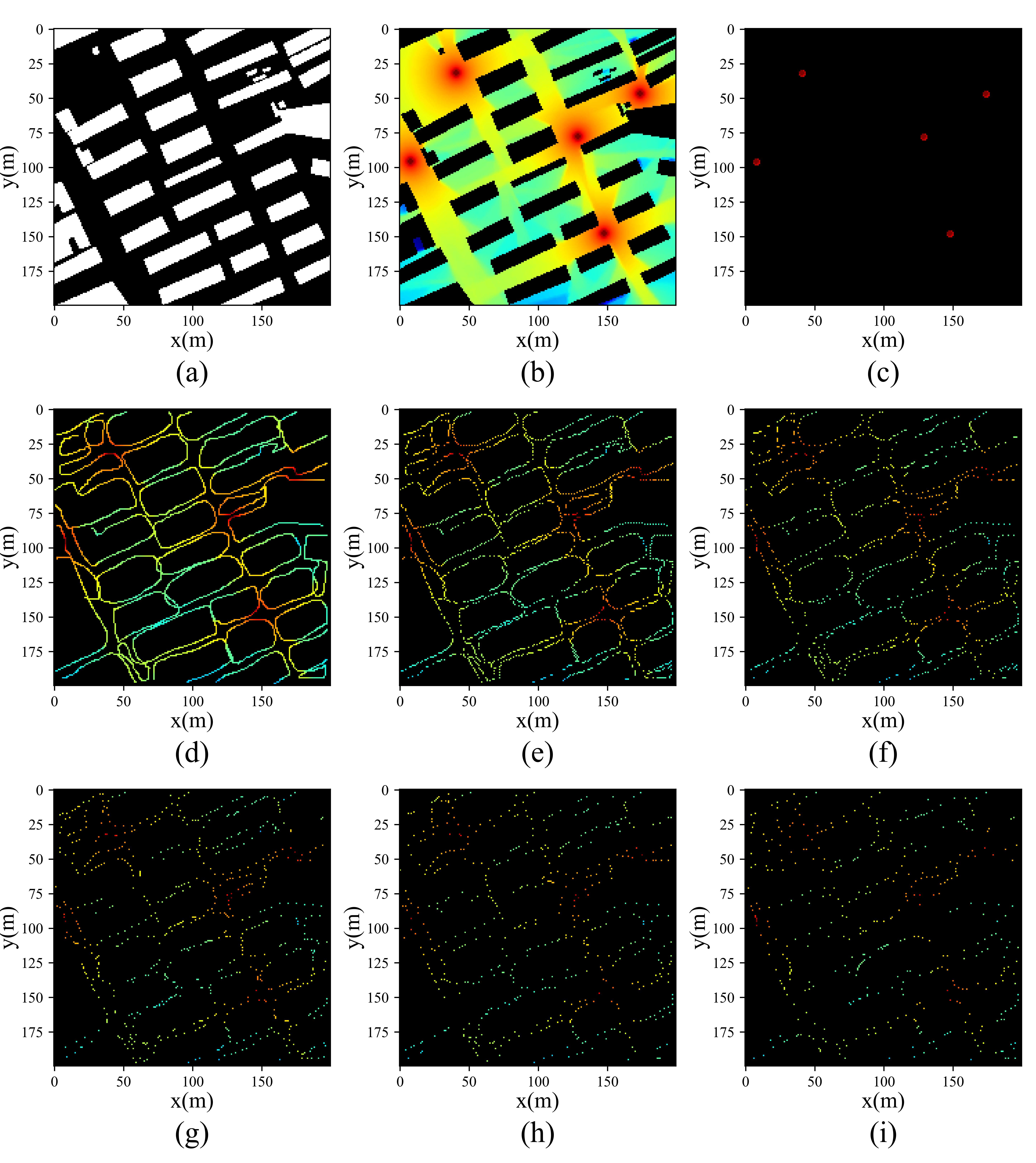}
		\caption{An example of building layouts, global radio maps, and local radio maps, along with their corresponding sampled maps at different sampling time intervals. (a) Environment map. (b) Global radio map. (c) Local radio map. (d-i) Sampled maps at intervals of 1, 2, 4, 6, 8, and 10 s.}
		\label{example}
		\vspace{-0.5cm}
	\end{figure}

	Therefore, our aim is not to propose a sampling strategy superior to uniform random sampling, but rather to simulate sampling conditions that are more representative of real-world scenarios. Specifically, we assume that a vehicle-mounted terminal performs measurements around buildings at a constant speed of 1 m/s. To simulate different sampling conditions, we use six different sampling time intervals for each environment, specifically 1, 2, 4, 6, 8, and 10 seconds (s). It is important to note that as the sampling time interval increases, the distance between adjacent sampling points becomes larger, resulting in lower sampling density. A schematic of this process is shown in Fig. \ref{example}.

	In the simulation experiments, the sampled maps are generated by resampling from the global radio map. Each sample contains six different sampling time intervals, meaning each sample can generate six versions of the sampled map, each with a different sampling density. Thus, single-source and multi-source global/local radio map subsets correspond to 34,560 sampled map samples.
	\begin{table}[!t]
		\centering
		\caption{VaryTxLoc Dataset parameters\label{tab:table1}}
		
		\label{para}
		\begin{tabular}{c c}
			\hline
			
			\rowcolor{gray!30}\textbf{Parameters} & \textbf{Value}\\
			\hline
			Map Size & 200 $\times$ 200 pixels\\
			
			Pixel length & 1.0 m\\
			
			Source,Receiver height & 1.5 m\\
			
			Building height & 25.0 m\\
			
			Center carrier frequnency & 2.0 GHz\\
			
			Bandwidth & 10.0 MHz\\
			
			Transmit power & 24.0 dBm\\
			
			Antenna type & Omnidirectional antenna\\
			
			Antenna gain & 10.0 dBi\\
			\hline
		\end{tabular}
		\vspace{-0.3cm}
	\end{table}
	
	6) \textit{Dataset Split:} 
	In the multi-source separation stage, the corresponding ground truth is the multi-source local radio map subset, while in the numerical coordinate regression stage, the corresponding ground truth is the single-source local radio map subset. Under the settings of 6 different sampling time intervals, the training sets for both stages consist of data from 18 building layouts, totaling 25,920 samples, while the validation and test sets each contain data from 3 building layouts, with 4,320 samples. It is important to note that the building layouts in the training, validation, and test sets are different to validate the applicability of the proposed method to new building layouts.
	
	7) \textit{OOD Test Dataset:} 
	In addition to the primary dataset, we also include an OOD test dataset to evaluate the robustness of the models. This OOD dataset consists of two types: one type includes samples with 10 sources per map, exceeding the maximum number of sources, 7, in the training samples. The other type contains samples with 3 sources, where the emission powers of the three sources are fixed at 12 dBm, 20 dBm, and 24 dBm. Each type of OOD data contains 3 environment maps, with 180 samples.
	
	8) \textit{Large-AoI Test Dataset:}
	The VaryTxLoc Dataset also include a Large-AoI test dataset to evaluate model performance on larger AoIs. This dataset consists of three AoI sizes: \( 400 \times 400 \, \text{m}^2 \), \( 600 \times 600 \, \text{m}^2 \), and \( 800 \times 800 \, \text{m}^2 \). For each AoI, samples with 1, 3, 5, and 7 sources were generated. Each source count has 20 different placement layouts under 6 sampling intervals, resulting in 480 samples per AoI and a total of 1,920 sources.

	\subsection{Performance Metrics}
	\label{section:B}
	The performance of the proposed approach and benchmarks is assessed using the following metrics, which follow the standard evaluation practices commonly adopted in the source localization research \cite{zhan2021deepmtl}, \cite{mitchell2022deep}, \cite{10246277}.
	\begin{enumerate}
		\item{Mean Localization Error (mLE):}
		mLE represents the average Euclidean distance between predicted and true coordinates of all sources.
		For each source \(s\), the localization error is defined as
		\begin{equation} 
			\epsilon_d=\sqrt{(x_{pc}^s-x_{tc}^s)^2 + (y_{pc}^s-y_{tc}^s)^2},
		\end{equation}
		where \((x_{tc}^s, y_{tc}^s)\) and \((x_{pc}^s, y_{pc}^s)\) denote the true and predicted coordinates of source \(s\), respectively.
		Lower mLE values reflect higher localization accuracy.
		\item{False Alarm Rate ($\epsilon_{far}$) and Missed Detection Rate ($\epsilon_{mdr}$):} 
		These metrics evaluate the accuracy of source localization predictions by comparing the predicted number of sources \(\hat{M}\) with the actual number of sources \(M\). 
		\begin{itemize}
			\item $\epsilon_{far}$ is defined as the ratio of excess predicted sources (i.e., the number of false positives) to the total number of predicted sources, capturing the frequency of over-predictions.
			
			\item $\epsilon_{mdr}$ is defined as the ratio of missed true sources (i.e., the number of false negatives) to the total number of actual sources, indicating the proportion of under-predictions.

		\end{itemize}
		Lower values of both metrics indicate better model performance, with an ideal outcome yielding $\epsilon_{far}=0$ and $\epsilon_{mdr}=0$.

		\item{OSPA metric:} OSPA provides a comprehensive measure that combines both localization and cardinality errors into a single value \cite{schuhmacher2008consistent}, making it particularly suitable for evaluating the performance of source localization methods. It penalizes not only errors in the estimated source positions but also discrepancies in the number of predicted sources. Lower OSPA values indicate better performance. A distance parameter \( g \) (in meters) is incorporated to limit the impact of large localization errors, while false alarms (over-predicted sources) and missed detections (under-predicted sources) are also penalized. When the number of true sources (\( M \)) differs from the number of estimated sources (\( \hat{M} \)), an error of \( g^2 \cdot |\hat{M} - M| \) is added to the total error.
		
		The OSPA distance is computed using the following formula when \( M \leq \hat{M} \):
		
		\begin{equation} 
			OSPA = \left(\frac{1}{\hat{M}} \sum_{s=1}^{\hat{M}}\epsilon_d^g(\hat{Q_s},Q_s)^2+g^2(\hat{M}-M)\right)^{\frac{1}{2}},
		\end{equation}
		
		where \(\epsilon_d^g(\hat{Q_s}, Q_s) = \min(g, \epsilon_d(\hat{Q_s}, Q_s))\). The parameter \( g \) controls the balance between penalizing false alarms and missed detections. In this work, we set \( g = 20 \).
		If the predicted number of sources is smaller than the true number (\( \hat{M} < M \)), the roles of \( M \) and \( \hat{M} \) in the formula are swapped.
		
	\end{enumerate}

	\subsection{Implementations}
	
	The implementation was done using Python version 3.9.12 and PyTorch \cite{paszke2019pytorch}, version 2.0.1.
	We conducted supervised learning on the VaryTxLoc dataset using the AdamW optimizer. The initial learning rate is set to $10^{-3}$ with a weight decay of $10^{-4}$. The learning rate is reduced by a factor of 10 after 10 epochs. Training is performed over 50 epochs with a batch size of 32. To prevent overfitting, we select the model parameters that achieve the lowest validation error during the 50 training epochs. 
	
	All experiments with CPU implementations are run on an Intel Core i5-12490F, while conducting all model training experiments on a GPU server with 16GB memory and Nvidia GeForce RTX 4060Ti GPU.

	\subsection{Comparisons with SOTA Methods}
	
	In this section, we present a comparative evaluation of the proposed method against three state-of-the-art (SOTA) methods, \textit{TL;DL} \cite{mitchell2022deep}, \textit{cGANPeak} \cite{10246277}, and \textit{IMNet} \cite{10829585}, as well as two additional baselines, \textit{KrigingPeak} and \textit{LRMPeak}.
	
	\textbf{\emph{TL;DL} \cite{mitchell2022deep}}: \textit{TL;DL} first employs a UNet to transform the raw measurement data into a heatmap, where each source location is represented by a \(3\times3\) pixel patch (center value 1, surrounding 0.5). Candidate sources are then identified via thresholding. However, this method disregards the actual RSS distribution surrounding each source. 
	
	\textbf{\emph{cGANPeak}\footnote{The dataset used in this work contains only 25,920 training samples, significantly smaller than the 100,000 samples used in \cite{10246277}, and our urban scenario is more complex. It is well known that cGANs are sensitive to the dataset during training, which may lead to a certain degradation in performance.} \cite{10246277}}: \textit{cGANPeak} uses a cGAN to convert raw measurement data into a global radio map, followed by persistence calculation and thresholding to identify peaks. Source positions are determined by computing the weighted sum of the four-pixel neighborhood. This approach includes redundant RSS information from distant areas, which dilutes attention to the critical RSS patterns around the sources.
	
	\textbf{\emph{IMNet} \cite{10829585}}: \textit{IMNet} reconstructs a global radio map from raw measurement data and then applies a two-dimensional constant false alarm rate (2D-CFAR) detector to identify peaks corresponding to source locations. Similar to \textit{cGANPeak}, it incorporates redundant RSS information from distant areas, which weakens the emphasis on key RSS features near the sources.
	
	\textbf{\emph{KrigingPeak}}: \textit{KrigingPeak} reconstructs a global radio map from raw measurements data using kriging interpolation, followed by the same heuristic localization strategy as in \cite{10246277}.
	
	\textbf{\emph{LRMPeak}}: \textit{LRMPeak} employs the proposed M2SNet to generate a local radio map and then applies the same heuristic localization procedure as in \cite{10246277}.

	\begin{table}[t]
		\centering
		\captionsetup{justification=centering, labelsep=newline} 
		\caption{Comparison of mLE, $\epsilon_{far}$, $\epsilon_{mdr}$, and OSPA for different methods. Bold values highlight the best performance. The $\downarrow$ indicates metrics whereby lower values constitute better outcomes.}
		\renewcommand\arraystretch{1.1}\resizebox{\columnwidth}{!}{
			\begin{tabular}{c|cccc}
				\toprule[1.5pt]
				\toprule \multirow{2}{*}{\textbf{Method}} & \multicolumn{4}{c}{\textbf{Metric}} \\
				\cline{2-5}
				
				&  \textbf{mLE (m) $\downarrow$} & \textbf{$\bm{\epsilon_{far} (\%) \downarrow}$} & \textbf{$\bm{\epsilon_{mdr} (\%) \downarrow}$} & \textbf{OSPA (m) $\downarrow$} \\
				
				\midrule
				KrigingPeak  & 3.959  & 11.078 & 17.500 &  17.081\\
				IMNet & 3.065&3.639&8.056&10.161 \\
				cGANPeak 	&1.363	&0.910	&9.580	&  9.748\\
				
				TL;DL 	&0.465	&3.333	& 0.208& 3.301 \\
				LRMPeak	&0.529	&\textbf{0.208}	&\textbf{0.035} & 0.603 \\
				LRM-MSL (Ours)  &\textbf{0.213}   & \textbf{0.208} & \textbf{0.035} &  \textbf{0.459} \\

				\bottomrule
				\bottomrule[1.5pt]
		\end{tabular}}
		\label{total}
	\end{table}
	
	\begin{table}[t!]
		\centering
		\captionsetup{justification=centering, labelsep=newline}
		\caption{Comparison of mLE, $\epsilon_{far}$, $\epsilon_{mdr}$, and OSPA for Different Methods across Varying Source Numbers. Bold values indicate the best performance.}
		\renewcommand\arraystretch{1.2}
		\resizebox{\columnwidth}{!}{
			\begin{tabular}{c|cccc|cccc}
				\toprule[1.5pt]
				\toprule
				\multirow{2}{*}{\textbf{Method}} & \multicolumn{4}{c|}{\textbf{1Tx}} & \multicolumn{4}{c}{\textbf{3Tx}} \\
				\cline{2-9}
				& \textbf{mLE (m)} & \textbf{$\bm{\epsilon_{far} (\%)}$} & \textbf{$\bm{\epsilon_{mdr} (\%)}$} & \textbf{OSPA (m)}
				& \textbf{mLE (m)} & \textbf{$\bm{\epsilon_{far} (\%)}$} & \textbf{$\bm{\epsilon_{mdr} (\%)}$} & \textbf{OSPA (m)} \\
				\midrule
				KrigingPeak      & 3.857 & 11.051 & 6.476 & 11.879 & 5.391 & 18.002 & 9.333& 14.971 \\
				IMNet            & 3.205 & 9.238 & 5.524 & 6.082 & 3.326 & 3.300 & 5.143 & 6.413\\
				cGANPeak         & 1.917 & 19.810 & 2.952 &9.994  & 1.911 & 1.714 & 10.190& 7.467 \\
				TL;DL            &0.452 & 0.667 & 0.667 &2.482 & 0.362 & 1.048 & 0.381 & 2.501 \\
				LRMPeak         & 0.503 & \textbf{0.000} & \textbf{0.000} & 0.503& 0.499 & \textbf{0.000} & 0.095 & 0.525 \\
				LRM-MSL (Ours)   & \textbf{0.229} & 0.190 & \textbf{0.000}  & \textbf{0.267} & \textbf{0.208} & 0.190 & \textbf{0.000}& \textbf{0.321}\\
				\bottomrule[0.8pt]
			\end{tabular}
		}
		\vspace{3pt} 
		\resizebox{\columnwidth}{!}{
			\begin{tabular}{c|cccc|cccc}
				\toprule[0.8pt]
				\multirow{2}{*}{\textbf{Method}} & \multicolumn{4}{c|}{\textbf{5Tx}} & \multicolumn{4}{c}{\textbf{7Tx}} \\
				\cline{2-9}
				& \textbf{mLE (m)} & \textbf{$\bm{\epsilon_{far} (\%)}$} & \textbf{$\bm{\epsilon_{mdr} (\%)}$} & \textbf{OSPA (m)}
				& \textbf{mLE (m)} & \textbf{$\bm{\epsilon_{far} (\%)}$} & \textbf{$\bm{\epsilon_{mdr} (\%)}$} & \textbf{OSPA (m)} \\
				\midrule
				KrigingPeak      & 4.437 & 3.326 & 16.952& 16.621 & 3.824 & 0.988 & 23.619 & 24.187\\
				IMNet            & 3.039 & 1.293 & 5.714 & 5.612  & 3.127 & 0.417 & 11.619 & 7.550 \\
				cGANPeak         & 1.794 & 1.524 & 11.810 & 7.760 & 1.599 & 0.381 & 16.571& 8.708\\
				TL;DL            &0.485 & 0.952 & 1.048 & 3.004  &0.499 & 1.524 & 0.667 & 3.218\\
				LRMPeak         & 0.531 & \textbf{0.000} & 0.571 & 0.642 & 0.387 & \textbf{0.000} & \textbf{0.095} & 0.454\\
				LRM-MSL (Ours)   & \textbf{0.203} & 0.285&\textbf{0.000}  &\textbf{0.486} & \textbf{0.213} & \textbf{0.000} & \textbf{0.095} & \textbf{0.345} \\
				\bottomrule
				\bottomrule[1.5pt]
			\end{tabular}
		}
		\label{varysource}
		\vspace{-0.5cm}
	\end{table}
	
	\begin{table}[t]
		\centering
		\captionsetup{justification=centering, labelsep=newline} 
		\caption{Comparison of mLE, $\epsilon_{far}$, $\epsilon_{mdr}$, and OSPA for Different Methods across varying Sampling Time Intervals. Larger intervals indicate sparser measurements. Bold values indicate the best performance.}
		\renewcommand\arraystretch{1.1}\resizebox{\columnwidth}{!}{
			\begin{tabular}{c|c|cccccc}
				\toprule[1.5pt]
				\toprule
				\multirow{2}{*}{\textbf{Metric}} & \multirow{2}{*}{\textbf{Method}} & \multicolumn{6}{c}{ \textbf{Sampling Time Interval (s)}} \\
				
				\cline{3-8}
				&  &  \textbf{1} & \textbf{2} & \textbf{4} & \textbf{6} &\textbf{8} &\textbf{10} \\
				\midrule
				\multirow{6}{*}{mLE (m) } & KrigingPeak  & 2.091  & 2.407  & 4.482    &  4.540  & 4.604  & 6.269\\
				\multirow{6}{*}{} & IMNet	& 2.610 & 2.570 & 2.925 & 3.188 & 3.327 & 3.829
				\\
				\multirow{6}{*}{} & cGANPeak &1.216&1.318&1.241&1.405&1.494  &   1.587	\\
				\multirow{6}{*}{} & TL;DL	&0.156 &0.249&0.419 &0.583  &0.649  &0.906\\
				\multirow{6}{*}{} & LRMPeak	&0.186&0.209&0.239&0.599&0.933&0.942	\\
				\multirow{6}{*}{} &   LRM-MSL (Ours) &\textbf{0.104}&\textbf{0.116}&\textbf{0.137}&\textbf{0.329}&\textbf{0.260}& \textbf{0.334}\\

				\midrule
				\multirow{6}{*}{$\bm{\epsilon$ $_{far}}$ (\%)} & KrigingPeak	& 3.319	& 4.867	& 9.231 & 13.111 & 14.750 & 21.382	\\
				\multirow{6}{*}{} & IMNet	& 0.430 & 0.680 & 2.870 & 5.858 & 4.110 & 7.611
				\\
				\multirow{6}{*}{} & cGANPeak	& 0.435	& 0.441	&\textbf{0.000}& 1.766  & 2.138 & 0.741	\\
				\multirow{6}{*}{} & TL;DL	& 0.208	& 0.417	& \textbf{0.000}	& 3.333 & 3.333 & 8.125	\\
				\multirow{65}{*}{} & LRMPeak	& \textbf{0.000}	& \textbf{0.208}	& \textbf{0.000}	& \textbf{0.208} & 0.208 & \textbf{0.621}	\\
				\multirow{6}{*}{} &  LRM-MSL (Ours)   & \textbf{0.000} & \textbf{0.208} & \textbf{0.000} & 0.415& \textbf{0.000} & \textbf{0.621} \\

				\midrule
				\multirow{6}{*}{$\bm{\epsilon$ $_{mdr}}$ (\%)} & KrigingPeak	& 8.958	& 10.417 & 13.958 & 18.542 & 28.958 & 24.167	\\
				\multirow{6}{*}{} & IMNet	& 3.542 & 8.750 & 8.333 & 6.250 & 12.500 & 8.958
				\\
				\multirow{6}{*}{} & cGANPeak	&4.583	&5.833	&9.375	& 7.292 & 14.167 & 16.250	\\
				\multirow{6}{*}{} & TL;DL & \textbf{0.000}	& \textbf{0.000}	& 0.208	& 0.208 & \textbf{0.000} & \textbf{0.000}	\\
				\multirow{6}{*}{} & LRMPeak	& \textbf{0.000}	& \textbf{0.000}	& \textbf{0.000}	& \textbf{0.000} & 0.208 & \textbf{0.000}	\\
				\multirow{6}{*}{} & LRM-MSL (Ours)   &\textbf{0.000} & \textbf{0.000}&\textbf{0.000} &\textbf{0.000} &\textbf{0.000} & 0.208 \\
				
				\midrule
				
				\multirow{6}{*}{OSPA (m)} & KrigingPeak  & 10.157 & 11.762 & 15.490 & 18.485 & 22.693 & 23.897 \\
				\multirow{6}{*}{} & IMNet & 5.314 & 8.653 & 9.664 & 10.496 & 13.245 & 13.596
				\\
				\multirow{6}{*}{} & cGANPeak &6.039&7.443&7.505&11.571&12.256&13.680	\\
				\multirow{6}{*}{} & TL;DL &0.857&1.145&1.463&3.958&5.415&6.968	\\
				\multirow{6}{*}{} & LRMPeak	 &0.155&	0.395&0.359&0.789&0.748&1.389	\\
				\multirow{6}{*}{} & LRM-MSL (Ours)  &\textbf{0.135}&\textbf{0.315}&\textbf{0.182}&\textbf{0.658}&\textbf{0.366}&\textbf{1.103}\\

				\bottomrule
				\bottomrule[1.5pt]
			\end{tabular}
		}
		\label{all_metric_interval}
		
	\end{table}

	\subsubsection{Overall Performance}
	Table~\ref{total} compares the proposed \textit{LRM-MSL} with baseline methods in terms of mLE, $\epsilon_{far}$, $\epsilon_{mdr}$, and OSPA.
	Overall, \textit{LRM-MSL} achieves the best performance across all metrics.
	\textit{TL;DL} overlooks the real RSS distribution around sources, while its performance is acceptable, it remains inferior to \textit{LRM-MSL}.
	\textit{KrigingPeak}, \textit{cGANPeak}, and \textit{IMNet} show poor capability to localize sources, the reason is that they all reconstruct a global radio map. For the same sparse measurements, reconstructing an accurate global radio map is inherently more challenging than reconstructing a local radio map.
	Moreover, the global radio map contains redundant RSS information from areas far from the sources and lacks a clear boundary between local and non-local areas.
	When combined with heuristic localization methods, these characteristics necessitate strict threshold tuning to detect potential source positions, and the complex RSS patterns in urban environments further degrade accuracy.
	In contrast, \textit{LRMPeak} also employs a heuristic localization strategy but benefits from the local radio map representation, which simplifies threshold selection and enhances localization accuracy.

	\subsubsection{Performance with Varying Source Numbers} 
	In this section, we fix the total number of sources in each setting at 1050 to ensure a fair comparison across varying source numbers. As shown in Table \ref{varysource}, our proposed \textit{LRM-MSL} method consistently achieves the best overall performance, demonstrating superior localization accuracy (mLE) as well as low $\epsilon_{far}$ and $\epsilon_{mdr}$. The \textit{LRMPeak} method also attains low $\epsilon_{far}$ and $\epsilon_{mdr}$, which benefits from the clear and sharp drop in pixel values between local and non-local areas inherent in the local radio map, but its mLE and OSPA are slightly inferior to those of \textit{LRM-MSL}. The \textit{TL;DL} method shows lower performance than \textit{LRM-MSL}, which may be attributed to its lack of utilizing the true RSS distribution during localization.
	
	In contrast, \textit{KrigingPeak}, \textit{IMNet}, and \textit{cGANPeak} exhibit less stable performance across varying source numbers. This instability primarily stems from their reliance on heuristic threshold-based localization on global radio maps. Due to the complex and highly variable electromagnetic propagation characteristics in urban environments, the optimal threshold varies significantly across individual samples. Since it is impractical to determine an optimal threshold for each sample, applying a fixed threshold inevitably leads to fluctuations in $\epsilon_{far}$ and $\epsilon_{mdr}$.
	This observation further highlights the advantage of the local radio map design, whose stable and well-defined boundaries between local and non-local areas enable \textit{LRM-MSL} to maintain robust performance under varying source numbers.

	\subsubsection{Performance with Varying Sampling Time Interval} 
	Table \ref{all_metric_interval} reports the localization performance under different sampling time intervals, where larger intervals correspond to sparser RSS measurements. As the interval increases, the reduced measurement density inevitably causes varying degrees of performance degradation across all methods.
	\textit{LRM-MSL} demonstrates remarkable robustness to increasing sampling intervals, consistently maintaining low mLE, $\epsilon_{far}$, $\epsilon_{mdr}$, and OSPA values even under the sparsest sampling conditions. \textit{LRMPeak} also sustains low $\epsilon_{far}$ and $\epsilon_{mdr}$ across intervals, however, its localization accuracy slightly lags behind that of \textit{LRM-MSL}.
	In contrast, \textit{TL;DL} maintains a low $\epsilon_{mdr}$ but suffers from elevated $\epsilon_{far}$ at larger intervals. This is likely because its reliance on $3 \times 3$ pixel patches provides limited gradient guidance for heatmap reconstruction, resulting in lower reconstruction accuracy, which becomes more pronounced under sparse sampling. Meanwhile, the performance of \textit{KrigingPeak}, \textit{IMNet}, and \textit{cGANPeak} deteriorates significantly as the sampling interval increases. Among them, \textit{KrigingPeak} performs the worst overall, primarily due to its interpolation-based global radio map reconstruction lacking adaptability and learning capability in complex urban environments.
	Overall, RSS information is critical for localization, and the sparsity of RSS measurements naturally impacts localization performance. The \textit{LRM-MSL} consistently demonstrates strong competitiveness across varying sampling intervals.

	\begin{table*}[ht]
		\centering
		\captionsetup{justification=centering, labelsep=newline} 
		\caption{Performance Comparison of Different Radio Map Reconstruction Models, Local Area Radii $r$, and Threshold Values $\gamma$ within the Proposed LRM-MSL Framework.}
		
		\renewcommand{\arraystretch}{1.1}
		\resizebox{\textwidth}{!}{
			\begin{tabular}{c | c | cccc | cccc | cccc}
				\toprule[1.5pt]
				\toprule
				\multirow{2}{*}{\textbf{Method}} & \multirow{2}{*}{$r$} 
				& \multicolumn{4}{c|}{\textbf{$\gamma=100$}} 
				& \multicolumn{4}{c|}{\textbf{$\gamma=127$}} 
				& \multicolumn{4}{c}{\textbf{$\gamma=150$}} \\
				\cmidrule(lr){3-6} \cmidrule(lr){7-10} \cmidrule(lr){11-14}
				& & $\epsilon_{\mathrm{far}} \downarrow$ & $\epsilon_{\mathrm{mdr}} \downarrow$ & mLE (m) $\downarrow$ & OSPA (m) $\downarrow$
				& $\epsilon_{\mathrm{far}} \downarrow$ & $\epsilon_{\mathrm{mdr}} \downarrow$ & mLE (m) $\downarrow$ & OSPA (m) $\downarrow$
				& $\epsilon_{\mathrm{far}} \downarrow$ & $\epsilon_{\mathrm{mdr}} \downarrow$ & mLE (m) $\downarrow$ & OSPA (m) $\downarrow$ \\
				\midrule
				PMNet        &  \multirow{4}{*}{1}     &  1.641  &  0.440  &  0.756  & 2.322   &  1.259  & 1.071   & 0.799   &  2.546  &  0.996  &  2.245  &  0.873 & 3.241  \\
				
				RadioUNet    &      &  \textbf{0.465} & 0.845   &  0.614  &  1.511 & \textbf{0.118}   & 1.823  &   0.621 &  2.029   &  \textbf{0.095}  & 3.096  & 0.653   & 2.886  \\
				RME-GAN        & & 0.810 & 0.723 & 0.423 & 1.580  &0.427& 1.406 & 0.437 & 1.808 & 0.278 & 2.269 & 0.442  & 2.329 \\
				M2SNet (Ours) &     & 0.896 &  \textbf{0.174} & \textbf{0.208}& \textbf{ 1.049} & 0.561 & \textbf{0.411}  & \textbf{0.209} & \textbf{0.965} & 0.343  & \textbf{0.932} & \textbf{0.208} & \textbf{1.186} \\
				\midrule
				
				PMNet         &   \multirow{4}{*}{2}   & 0.364 & 0.064  &  0.574 &  0.880  &  0.449 &  0.058  &  0.558  & 0.929 & 0.346  & 0.098   &  0.614  &0.937  \\
				RadioUNet    &       &  0.559  & 0.133   &  0.437  & 0.959 & 0.364 & 0.203   & 0.445   & 0.878 &  0.255  & 0.295   &  0.454  & 0.881  \\
				RME-GAN         &  & 0.913 & 0.185 & 0.391 & 1.224  & 0.674& 0.226 & 0.391 & 1.094 & 0.612 & 0.301 & 0.395 & 1.091 \\
				M2SNet (Ours) &      & \textbf{0.237} & \textbf{0.035} & \textbf{0.212}& \textbf{0.484} & \textbf{0.208} & \textbf{0.035} & \textbf{0.213} & \textbf{0.459} &\textbf{0.208} & \textbf{0.046} & \textbf{0.218} & \textbf{0.472}  \\
				\midrule
				
				PMNet         &   \multirow{4}{*}{3}  & 0.576 & 0.023 &0.601 & 1.060  &0.404&0.029 &0.606& 0.935  & 0.358 & 0.041  & 0.634& 0.941 \\
				RadioUNet  &   &\textbf{0.104} &\textbf{0.017}&0.414& \textbf{0.505} &\textbf{0.093}&\textbf{0.006}&0.401&\textbf{0.457} &\textbf{0.127}&\textbf{0.017}&0.404&\textbf{0.495} \\
				RME-GAN        &  & 0.644 & 0.029 & 0.393 & 0.879 & 0.569 & 0.029 & 0.394 & 0.829 & 0.461 & 0.041 & 0.394 & 0.775\\
				M2SNet (Ours) &     & 0.415 & 0.029 & \textbf{0.276} & 0.614 & 0.415 & 0.035 & \textbf{0.277} & 0.599  & 0.306 & 0.035 & \textbf{0.276} & 0.529 \\
				\bottomrule
				\bottomrule[1.5pt]
				\multicolumn{14}{p{21cm}}{
					Since most background pixels in the local radio map are zero, averaging over all pixels dilutes local differences, making RMSE less discriminative. The False Alarm Rate ($\epsilon_{\mathrm{far}}$) and Missed Detection Rate ($\epsilon_{\mathrm{mdr}}$) are more sensitive to source region accuracy and thus provide more meaningful evaluation.
				} \\

		\end{tabular}}
		
		\label{tab:sr_performance}
		\vspace{-0.5cm}
	\end{table*}

	\subsection{Ablation Study}
	
	\subsubsection{Ablation Study on Reconstruction Models and Parameters}
	In this section, we conduct an ablation study to analyze the impact of different radio map reconstruction models (PMNet~\cite{lee2023pmnet}, RadioUNet~\cite{9354041}, and RME-GAN~\cite{10130091}), local area radii $r$, and threshold values $\gamma$ used in the connected component analysis on the performance of the proposed \textit{LRM-MSL} framework, where the second stage localization is performed using SourceNet. As shown in Table \ref{tab:sr_performance}, M2SNet achieves the best or near-best performance across most configurations, while also exhibiting a significant advantage in computational efficiency, as reported in Table \ref{step1efficiency}. 
	Although M2SNet shows slightly lower performance than RadioUNet when $r=3$, it achieves the smallest parameter count and the shortest inference time among all compared methods.
	
	Notably, when the local area radius is set to $r=1$ or $r=2$, the mLE remains small. However, for $r=1$, the extremely limited spatial coverage provides insufficient gradient guidance for accurate radio map reconstruction, resulting in degraded reconstruction quality and a significantly higher $\epsilon_{\mathrm{mdr}}$.
	When $r=3$, the performance is comparable to that of $r=2$, but the spatial separation requirement between sources becomes more stringent, as the minimum allowable inter-source spacing increases to 6 m.
	
	Regarding the selection of the threshold $\gamma$, the reconstructed local radio maps exhibit a fixed and well-defined boundary between the local area and the non-local area, with a sharp change in pixel values. As discussed in Remark \ref{Remark1}, the pixel values within the local area change but remain within the range [200, 255]. Therefore, the choice of $\gamma$ is relatively flexible. In Table \ref{tab:sr_performance}, we evaluate three representative thresholds $\gamma=100,127,150 $ to assess the robustness of our method.
	The results show that, although mLE, false alarm rate, missed detection rate, and OSPA vary slightly across different $\gamma$ values, the overall performance remains stable, demonstrating the robustness of our method to the threshold setting. Since the pixel range is [0, 255], we adopt $\gamma = 127$ (i.e., half of the maximum pixel value) as the default threshold for the remaining experiments.

   	\begin{table}[t!]
   		\begin{center}
   			\centering
   			\captionsetup{justification=centering, labelsep=newline} 
   			\caption{Comparison of Computational Complexity and Inference Time for Different Radio Map Reconstruction Models.}
   			\label{step1efficiency}
   			\renewcommand\arraystretch{0.7}\resizebox{\columnwidth}{!}{
   				\begin{tabular}{c | c c c}  
   					\toprule[0.8pt]
   					\toprule[0.4pt]
   					\textbf{Method}  & \textbf{FLOPs (G)} & \textbf{params (M)} & \textbf{Inf Time (ms)}\\
   					\midrule
   					PMNet  & 10.22 & 8.00 & 5.01\\
   					RadioUNet& 36.11 & 31.04& 7.46\\
   					RME-GAN & 37.10 & 33.81  & 8.34\\
   					M2SNet (Ours) & \textbf{8.38} & \textbf{0.59} & \textbf{4.64}\\
   					
   					\bottomrule[0.4pt]
   					\bottomrule[0.8pt]
     			\end{tabular}}
   		\end{center}
   		\vspace{-0.2cm}
   	\end{table}

	\subsubsection{Ablation Study on Coordinate Estimation}
	
	\begin{table}[t!]
		\begin{center}
			\centering
			\captionsetup{justification=centering, labelsep=newline} 
			\caption{Comparison of Localization Accuracy, Computational Complexity, and Inference Time for Localization Methods.}
			\label{resnet}
			\renewcommand\arraystretch{1}\resizebox{\columnwidth}{!}{
				\begin{tabular}{c | c c c c}  
					\toprule[1pt]
					\toprule[0.5pt]
					\textbf{Method} & \textbf{mLE (m)} & \textbf{FLOPs (G)} & \textbf{params (M)} & \textbf{Inf Time (ms)}\\
					\midrule
					ArgMax & 1.928 & - & - & \textbf{0.20}\\
					Center-of-Mass & 0.613 & - & - & 0.40\\
					ResNet18 & 0.586 & \textbf{1.50} & 11.17 & 2.18\\
					ResNet34 & 0.575 & 2.83 & 21.28 & 3.48\\
					ResNet50 & 0.333 & 3.47 & 23.51 & 4.39\\
					SourceNet\ w/o\ CA & 0.436 & 2.82 & \textbf{9.78} & 2.33\\
					SourceNet (Ours) & \textbf{0.213} & 2.83 & 9.84 & 4.35\\
					\bottomrule[0.5pt]
					\bottomrule[1pt]
			\end{tabular}}
		\end{center}
		\vspace{-0.5cm}
	\end{table}

	In this section, we conduct an ablation study on the second stage of the \textit{LRM-MSL} framework, focusing on numerical coordinate estimation. We compare heuristic peak-finding methods (ArgMax and Center-of-Mass), classical deep learning models (ResNet variants), and our proposed SourceNet. All methods are evaluated on the single-source local radio maps $\hat{I}_{SS}^m$ generated by M2SNet with $\gamma=127$ and $r=2$.
	As shown in Table~\ref{resnet}, heuristic methods such as ArgMax and Center-of-Mass are extremely lightweight but yield poor localization accuracy, with mLEs of 1.928\,m and 0.613\,m, respectively. ResNet models improve accuracy but still fall short of SourceNet, while having substantially larger parameter counts, despite their lower FLOPs. In contrast, SourceNet achieves the best overall performance, attaining an mLE of only 0.213 m while keeping the model size small and inference time reasonable (4.35 ms). Such millisecond inference time is acceptable for real-time deployment.
	We also evaluate the effect of the CA module in SourceNet. Removing CA results in a notable accuracy drop with no significant computational savings, demonstrating the CA module’s importance in enhancing feature interaction for precise localization.
	Overall, SourceNet strikes a favorable balance between accuracy, parameter efficiency, and inference time.

	\subsection{Empirical Analysis of the Local Radio Map}
	
	To further validate the rationale of the local radio map design, we conduct an empirical analysis comparing localization performance using global and local radio maps. Specifically, we employ M2SNet to reconstruct the global radio map and local radio map ($r=1,2,3$). Since our SourceNet is designed for single-source localization after multi-source separation, each radio map used in this comparison contains only one source. These maps are then fed into SourceNet for localization. As shown in Table~\ref{tab:local_vs_global}, local radio maps consistently outperform the global radio map in terms of mLE. This confirms that RSS values outside the vicinity of sources are redundant for localization, and accurate position inference can be achieved using only the local RSS distributions. Although Table~\ref{tab:local_vs_global} shows that $r=1$ yields slightly better localization accuracy, as discussed earlier it suffers from higher false alarm and missed detection rates. Therefore, $r=2$ is adopted in this work for its overall superior trade-off.
	
	\begin{table}[t!]
		\renewcommand{\arraystretch}{1.1}
		\centering
		\captionsetup{justification=centering, labelsep=newline}
		\caption{Comparison of SourceNet Localization Accuracy Using Global Radio Maps and Local Radio Maps with Different $r$ Reconstructed by M2SNet
		\label{tab:local_vs_global}}
		
		\begin{tabular}{ccccc}
			\toprule[0.8pt]
			\toprule[0.4pt]
			\multirow{2}{*}{\textbf{Metric}} & \multirow{2}{*}{\textbf{Global Radio Map}} & \multicolumn{3}{c}{\textbf{Local Radio Map}} \\
			
			& & \textbf{r = 1} & \textbf{r = 2} & \textbf{r = 3} \\
			\hline
			mLE (m) & 0.686 & 0.217 & 0.252 & 0.301 \\
			\bottomrule[0.4pt]
			\bottomrule[0.8pt]
		\end{tabular}
	\end{table}

	\subsection{Robustness to Other Conditions}
	In this section, we examine the adaptability of the proposed method, along with all comparison methods, by analyzing their performance on noisy data and OOD data. By comparing the experimental results, we can gain a deeper understanding of the proposed method's performance when facing various challenges and demonstrate its robustness advantages.

	\begin{table}[t]
		\centering
		\captionsetup{justification=centering, labelsep=newline} 
		\caption{Comparison of mLE, $\epsilon_{far}$, $\epsilon_{mdr}$, and OSPA for Different Methods across Varying Noise Levels.}
		\renewcommand\arraystretch{1.1}\resizebox{\columnwidth}{!}{
			\begin{tabular}{c|c|cccc}
				\toprule[1.5pt]
				\toprule
				\multirow{2}{*}{\textbf{\thead{Noise \\ Level}}} & \multirow{2}{*}{\textbf{Method}} & \multicolumn{4}{c}{\textbf{Metric}} \\
				\cline{3-6}
				
				&  &  \textbf{mLE (m)} & \textbf{$\bm{\epsilon_{far} (\%)}$} & \textbf{$\bm{\epsilon_{mdr} (\%)}$} & \textbf{OSPA (m)} \\
				
				\midrule
				
				\multirow{4}{*}{\textbf{$\sigma=1dB$}} & IMNet&  3.048 & 3.920 & 9.792 & 11.178\\
				\multirow{4}{*}{} & cGANPeak 	&1.332	& 1.252	&9.721	&9.595\\
				\multirow{4}{*}{} & TL;DL	&0.520	&3.125	&\textbf{0.000}&3.997\\

				\multirow{4}{*}{} & LRM-MSL (Ours)  & \textbf{0.363} & \textbf{0.208}  &0.035 & \textbf{0.653}  \\
				
				\midrule
				\multirow{4}{*}{\textbf{$\sigma=3dB$}} & IMNet & 3.194 & 7.588 & 10.347 & 13.332 \\
				\multirow{4}{*}{} & cGANPeak	&1.847	&6.842	&7.816	&14.071	\\
				\multirow{4}{*}{} & TL;DL		&\textbf{0.869}	&3.750	&\textbf{0.000}	&4.968	\\
				\multirow{4}{*}{} & LRM-MSL (Ours)  &1.088  &\textbf{0.382}  &0.313	 & \textbf{1.786} \\
				
				\bottomrule
				\bottomrule[1.5pt]
		\end{tabular}}
		\label{noise}
		\vspace{-0.5cm}
	\end{table}

	\begin{table}[t]
		\centering
		\captionsetup{justification=centering, labelsep=newline} 
		\caption{Comparison of mLE, $\epsilon_{far}$, $\epsilon_{mdr}$, and OSPA for Different Methods across Varying OOD Conditions.}
		\renewcommand\arraystretch{1.1}\resizebox{\columnwidth}{!}{ 
			\begin{tabular}{c|c|cccc}
				\toprule[1.5pt]
				\toprule
				\multirow{2}{*}{\textbf{\thead{OOD \\ Condition}}} & \multirow{2}{*}{\textbf{Method}} & \multicolumn{4}{c}{\textbf{Metric}} \\
				\cline{3-6}
				
				&  &  \textbf{mLE (m)} & \textbf{$\bm{\epsilon_{far} (\%)}$} & \textbf{$\bm{\epsilon_{mdr} (\%)}$} & \textbf{OSPA (m)} \\
				\midrule
				\multirow{4}{*}{10Tx} & IMNet  & 3.152 & 5.808& 3.556 &10.230 \\
				\multirow{4}{*}{} &  cGANPeak	&2.002  &0.056& 17.667	& 7.460\\
				
				\multirow{4}{*}{} & TL;DL	&0.586	& 0.222&2.833& 3.687 \\

				\multirow{4}{*}{} & LRM-MSL (Ours)  & \textbf{0.218} & \textbf{0.056} & \textbf{0.333} & \textbf{1.135} \\

				\midrule
				\multirow{4}{*}{VaryPower} & IMNet    & 2.688& 9.639 & 6.667& 8.678\\
				\multirow{4}{*}{} & cGANPeak 	&2.282&3.333&3.148& 3.660\\
				
				\multirow{4}{*}{} & TL;DL	&0.930 & 4.444 & 4.630 & 6.275\\

				\multirow{4}{*}{} & LRM-MSL (Ours)  & \textbf{0.693} & \textbf{0.000}&\textbf{1.852}& \textbf{1.766} \\
				\bottomrule
				\bottomrule[1.5pt]
		\end{tabular}}
		\label{OOD}
		
	\end{table}

	\subsubsection{Results with Noisy Data} 
	To evaluate the robustness of our method under measurement noise, we follow a widely adopted practice in related work \cite{9523765}, \cite{wang2022mt}, \cite{10122907}, where additive Gaussian white noise is used to simulate the uncertainty in RSS measurements. Moreover, the use of Gaussian noise is theoretically supported by principles from differential entropy \cite{1055832} and the Central Limit Theorem (CLT) \cite{feller1971introduction}. Specifically, under a fixed variance constraint, the Gaussian distribution has the highest differential entropy, making it the most unpredictable noise model in the information-theoretic sense. In addition, the CLT states that the sum of a large number of statistically independent random variables tends to follow a normal distribution, which provides a solid probabilistic justification for approximating other types of aggregated random perturbations with a Gaussian distribution.
	Specifically, we incorporate zero-mean Gaussian noise with standard deviations of $\sigma = 1$ dB and $\sigma = 3$ dB to evaluate the method under different noise levels.
	
	As shown in Table \ref{noise}, the proposed method demonstrates competitive performance compared to \textit{IMNet}, \textit{TL;DL} and \textit{cGANPeak} across key metrics such as mLE, $\bm{\epsilon_{far}}$, $\bm{\epsilon_{mdr}}$, and OSPA.
	Specifically, for $\sigma = 0$ dB (clean) and $\sigma = 1$ dB, the proposed method achieves the lowest or close-to-lowest values across all key metrics, indicating strong robustness and overall performance under low to moderate noise. Under the noise condition of $\sigma = 3$ dB, while there is a slight increase in mLE and $\bm{\epsilon_{mdr}}$, the proposed method still maintains an edge in OSPA and $\bm{\epsilon_{far}}$.
	
	\subsubsection{Results with OOD Data} 
	In this section, we evaluate the performance of the proposed method using OOD test data, comparing it with \textit{IMNet}, \textit{TL;DL} and \textit{cGANPeak}. We consider two OOD scenarios: “10Tx”, where there are ten sources, and “VaryPower”, where the sources within the AoI operate with different transmission powers.
	The “10Tx” scenario represents cases where more sources (\(M^{\prime}\)) are present in the AoI than the number of sources (\(M\)) used during training. Such conditions are challenging for offline-trained systems, as retraining them is often impractical and time-consuming. The variability in the number of sources can occur unpredictably, highlighting the need for the model to adapt effectively without requiring extensive retraining.
	In real-world applications, sources may operate at different emission powers. In the "VaryPower" scenario, the three sources within the AoI operate at 12 dBm, 20 dBm, and 24 dBm, respectively. We evaluate the proposed method’s ability to adapt to such power variations. The results in Table \ref{OOD} demonstrate that the proposed method outperforms all comparison methods when handling OOD data.
	It consistently achieves the best results across both OOD scenarios, highlighting its robustness and adaptability to such conditions. This suggests that the proposed method is more effective in maintaining reliable localization performance in diverse and challenging environments.

	\begin{table}[t]
		\centering
		\captionsetup{justification=centering, labelsep=newline} 
		\caption{Comparison of mLE, $\epsilon_{far}$, $\epsilon_{mdr}$, and OSPA for Different Methods Applied to Larger AoIs via Pixel Scaling.}
		\renewcommand\arraystretch{1.1}\resizebox{\columnwidth}{!}{ 
			\begin{tabular}{c|c|cccc}
				\toprule[1.5pt]
				\toprule
				\multirow{2}{*}{\textbf{Scale}} & \multirow{2}{*}{\textbf{Method}} & \multicolumn{4}{c}{\textbf{Metric}} \\
				\cline{3-6}
				
				&  &  \textbf{mLE (m)} & \textbf{$\bm{\epsilon_{far} (\%)}$} & \textbf{$\bm{\epsilon_{mdr} (\%)}$} & \textbf{OSPA (m)} \\
				\midrule
				\multirow{4}{*}{$\times 2$ } & IMNet  & 4.971 & 19.844 & 8.281 &13.730 \\
				\multirow{4}{*}{} &  cGANPeak	& 5.897 & 0.677 & 22.292	& 10.904 \\
				
				\multirow{4}{*}{} & TL;DL	& 2.037	& 3.698 & 3.333 & 6.138 \\

				\multirow{4}{*}{} & LRM-MSL (Ours)  & \textbf{1.269} & \textbf{0.518} & \textbf{0.000} & \textbf{1.660} \\

				\midrule
				\multirow{4}{*}{$\times 3$ } & IMNet    &  5.027 & 18.229 & 12.708	& 13.761  \\
				\multirow{4}{*}{} & cGANPeak 	&5.087 & \textbf{0.000} & 17.448 & 9.548\\
				
				\multirow{4}{*}{} & TL;DL	&2.646 & 2.708 & 0.417 & 4.739 \\

				\multirow{4}{*}{} & LRM-MSL (Ours)  & \textbf{1.556} & 2.439 &\textbf{0.000}& \textbf{3.406} \\
				
				\midrule
				\multirow{4}{*}{$\times 4$ } & IMNet    & 6.100 & 6.719  & 26.302 & 14.444\\
				\multirow{4}{*}{} & cGANPeak 	&7.078 & \textbf{0.000} & 22.917 & 11.414\\
				 
				\multirow{4}{*}{} & TL;DL	& 3.488 & \textbf{0.000} & 2.604 & \textbf{5.072} \\

				\multirow{4}{*}{} & LRM-MSL (Ours)  & \textbf{2.577} & 4.478 &\textbf{0.000}& 5.875 \\
				\bottomrule
				\bottomrule[1.5pt]
		\end{tabular}}
		\label{LargeAoI}
		\vspace{-0.5cm}
	\end{table}

	\subsection{Performance Analysis under Larger AoI}
	In this section, we evaluate the performance of the proposed method on the Large-AoI test dataset, comparing it with \textit{IMNet}, \textit{TL;DL}, and \textit{cGANPeak}. The test dataset includes three AoI sizes: \( 400 \times 400 \, \text{m}^2 \), \( 600 \times 600 \, \text{m}^2 \), and \( 800 \times 800 \, \text{m}^2 \), which are larger than the \( 200 \times 200 \, \text{m}^2 \) areas used during training. To apply the pre-trained \textit{LRM-MSL} to these larger AoIs without modifying the network, all samples were resized to \( 200 \times 200 \) inputs via pixel scaling (2 m/pixel, 3 m/pixel, and 4 m/pixel, respectively). As shown in Table \ref{LargeAoI}, \textit{LRM-MSL} consistently outperforms the comparison methods across all scaling factors. Specifically, it achieves mLE values of 1.269 m, 1.556 m, and 2.577 m for scaling factors $\times 2$, $\times 3$, and $\times 4$, respectively. 
	It is observed that the localization performance of all methods decreases under larger AoIs, and this can be attributed to two main factors:
	First, \textit{LRM-MSL} achieves sub-pixel precision (mLE = 0.213 m) at the original 1 m/pixel resolution. However, when each pixel represents a larger physical area, even maintaining sub-pixel precision on the scaled input inevitably results in higher absolute localization errors.
	Second, the difference in pixel resolution between training and testing samples leads to a deviation in the sampled RSS distribution features between them, which may consequently degrade the model performance.
	Although the absolute mLE increases with the AoI size due to reduced spatial resolution, the localization accuracy of proposed \textit{LRM-MSL} remains competitive, supporting its applicability. Other metrics, including $\epsilon_{far}$, $\epsilon_{mdr}$, and OSPA, further demonstrate the superiority of \textit{LRM-MSL} over alternative methods.

	\subsection{Discussion of Failure Modes and Potential Solutions}
\begin{figure}[t]
		\centering
		\includegraphics[width=0.2\textwidth]{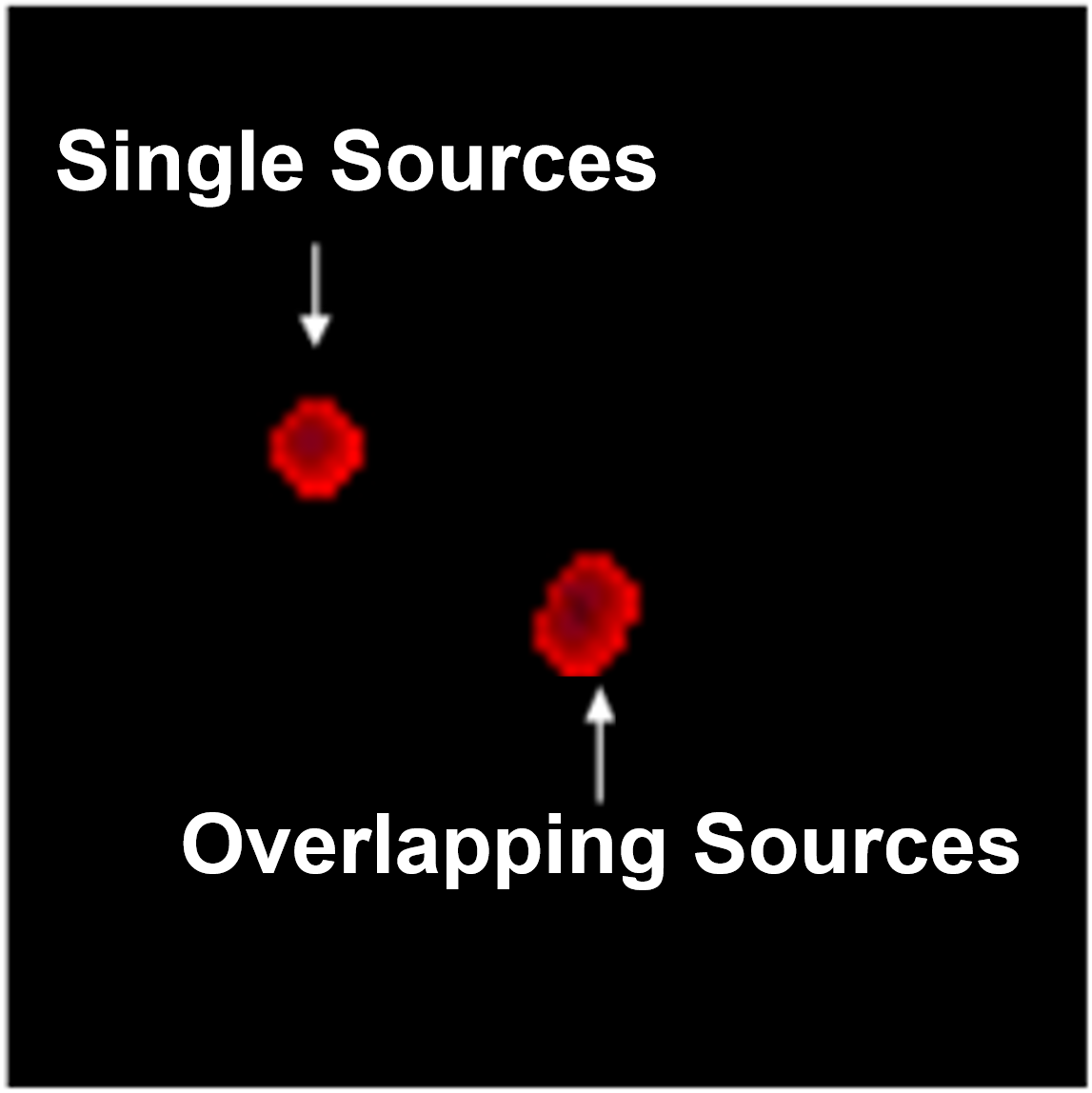}
		\caption{Example of merged connected components caused by overlapping local areas.}
		\label{overlap}
		\vspace{-0.5cm}
	\end{figure}

	Failure modes of the \textit{LRM-MSL} primarily occur when the inter-source spacing is less than 4 m. In such cases, the local areas of adjacent sources, designed as circles with a radius of 2 m, overlap, resulting in merged connected components that cause missed detections. These merged connected components often deviate from the expected circular shape and exhibit larger areas than those of individual sources. The overlapping situation of adjacent sources is illustrated in Fig. \ref{overlap}. To address this limitation, shape- and area-based post-processing techniques can be employed to identify and separate merged connected components.

	\section{Conclusion}  \label{Conclusion}
	This paper presents a Local Radio Map-Aided Multiple Source Localization Framework (LRM-MSL) as an innovative solution to MSL problem in complex urban environments. 
	The LRM-MSL is structured in a two-stage process. In the first stage, we employ M2SNet to construct the local radio map and then separate multiple sources, transforming the MSL problem into multiple SSL tasks.
	In the second stage, we design SourceNet for numerical coordinate regression to achieve precise localization of the single source. 
	Experimental results indicate that LRM-MSL outperforms state-of-the-art methods in various metrics, including mLE, $\epsilon_{far}$, $\epsilon_{mdr}$ and OSPA, while demonstrating generalization to an arbitrary number of sources and scalability to source numbers not included in the training dataset.
	In future work, we aim to extend the applicability of our framework to dense‑source scenarios with inter-source spacing below 4\,m, and expand our VaryTxLoc dataset to include real-world measurement data, with the aim of improving robustness in real-world deployments, and explore the use of super-resolution reconstruction techniques to enable accurate localization over even larger AoIs.

	\bibliographystyle{IEEEtran}
	\bibliography{IEEEabrv,reference}

	\vspace{11pt}

\end{document}